\documentclass[preprint]{aastex}
\usepackage[]{epsfig}
\usepackage[]{rotating}
\newcommand       \beq          {\begin{equation}}
\newcommand       \eeq          {\end{equation}}
\newcommand       \abs          {{\rm abs}}
\newcommand       \Angstrom     {\,{\rm \AA}}          
\newcommand	  \bT		{{\bf T}}
\newcommand	  \C		{{\rm C}}
\newcommand       \cm           {\,{\rm cm}}
\newcommand	  \tc	{^{\rm tc}}

\newcommand       \eV           {\,{\rm eV}\,}
\newcommand	  \g		{\,{\rm g}}
\newcommand	  \rmH		{{\rm H}}
\newcommand       \K            {\,{\rm K}}

\newcommand	  \INT		{{\rm int}}

\newcommand	  \rad		{{\rm rad}}

\newcommand	  \td		{^{\rm td}}

\newcommand       \simgt        {\gtrsim}
\newcommand       \gtsim        {\gtrsim}
\newcommand       \ltsim        {\lesssim}
\newcommand	  \Nb		{M}

\newcommand	  \figwidth	{12cm}
\newcommand	  \figwidthsmall	{11cm}
\countdef\decade=200
\advance\decade by \year
\countdef\hours=201
\advance\hours by \time
\divide\hours by 60
\countdef\mins=202
\advance\mins by \hours
\multiply\mins by 60
\multiply\hours by 100
\countdef\miltime=203
\advance\miltime by \hours
\advance\miltime by \time
\advance\miltime by -\mins

\begin{document}

\title{
	\vspace*{-2.0em}
	{\normalsize\rm submitted to {\it The Astrophysical Journal}}\\
	\vspace*{1.0em}
Infrared Emission from Interstellar Dust.\\
	I. Stochastic Heating of Small Grains\\
	}

\author{B.T. Draine and Aigen Li}
\affil{Princeton University Observatory, Peyton Hall,
        Princeton, NJ 08544, USA; 
	{\tt draine@astro.princeton.edu, agli@astro.princeton.edu}
	}

\begin{abstract}
We present a method for calculating the infrared emission from
a population of dust grains heated by starlight, including 
very small grains for which stochastic heating by starlight photons
results in high temperature transients.
Because state-to-state transition rates are generally unavailable
for complex molecules, we consider model PAH, graphitic, 
and silicate grains
with realistic vibrational mode spectra and realistic radiative properties.
The vibrational density of states is used in a statistical-mechanical 
description of the emission process.
Unlike previous treatments, our approach fully incorporates multiphoton
heating effects, important for large grains or strong radiation fields.

We discuss how the ``temperature'' of the grain is related to its
vibrational energy.
By comparing with an ``exact'' statistical calculation of the emission 
process, we determine the conditions under which the ``thermal'' and
the ``continuous cooling'' approximations
can be used to calculate the emission spectrum.

We present results for the infrared emission spectra of PAH grains of
various sizes heated by starlight.
We show how the relative strengths of the 6.2, 7.7,
and 11.3$\micron$ features depend on grain size, starlight spectrum
and intensity, and grain charging conditions.
We show results for 
grains in the ``cold neutral medium'', ``warm ionized medium'',
and representative conditions in photodissociation regions.
Our model results are compared to
observed ratios of emission features for the Milky Way and other
galaxies, and for the M17 and NGC 7023 photodissociation regions.

\end{abstract}


\section{Introduction}

Very small grains -- consisting of tens to hundreds of atoms -- play
a major role in the infrared emission from the interstellar medium.
These grains are small enough that the time-averaged vibrational energy
$\langle E\rangle$ is smaller than or comparable 
to the energy of the starlight photons
which heat the grains.
Stochastic heating by absorption of starlight therefore results in
transient ``temperature spikes'', during which much of the energy deposited
by the starlight photon is reradiated in the infrared.

The idea of transient heating of very small grains was first introduced 
by Greenberg (1968). Its importance was not apparent
until the detection of the near infrared emission of reflection nebulae 
(Sellgren et al. 1983) and detection by the
{\it Infrared Astronomical Satellite (IRAS)} of 12 and 25$\mu$m 
Galactic emission
which was far in excess of the emission expected for interstellar dust
at $T\approx20\K$
(Boulanger \& P\'erault 1988).
Subsequent measurements by the {\it Diffuse Infrared Background Experiment
(DIRBE)} instrument on the {\it Cosmic Background Explorer (COBE)} satellite
confirmed this and detected additional 
broadband emission at 3.5 and 4.9$\micron$
(Arendt et al.\ 1998).
More recently, spectrometers aboard the 
{\it Infrared Telescope in Space (IRTS)} 
(Onaka et al. 1996; Tanaka et al.
1996) and the {\it Infrared Space Observatory (ISO)} (Mattila et al. 1996) 
have shown
that the diffuse interstellar medium radiates strongly in emission
features at 3.3, 6.2, 7.7, 8.6, and 11.3$\micron$.
These emission features, previously
observed from a small number of 
bright reflection nebulae, planetary nebulae, and
HII regions, have been tentatively identified as emission from 
polycyclic aromatic hydrocarbons, or PAHs
(L\'eger \& Puget 1984; Allamandola, Tielens, \& Barker 1985).

To understand these observations we need to be able to calculate the
infrared emission expected for a given model of interstellar dust.
The question we address here is: what approximations can be used
to carry out accurate calculations of thermal emission 
from a population of very small grains exposed to stochastic heating by
starlight?

There have been a number of previous studies of the
discrete heating of very small grains
(Duley 1973;
Greenberg \& Hong 1974;
Gail \& Seldmayr 1975;
Greenberg 1976;
Purcell 1976;
Drapatz \& Michel 1977;
Andriesse 1978;
Aannestad \& Kenyon 1979;
L\'eger \& Puget 1984; 
Draine \& Anderson 1985; 
Puget et al. 1985;
D\'esert et al. 1986;
Dwek 1986; 
Guhathakurta \& Draine 1989;
Aannestad 1989;
Siebenmorgen et al. 1992;
Manske \& Henning 1998;
Duley \& Poole 1998).
Most studies have made three approximations: 
(1) the ``thermal'' approximation for excitation;
(2) use of ``bulk'' specific heats; and
(3) the ``continuous cooling'' approximation.
While accurate for large grains,
these approximations
could potentially introduce significant inaccuracies for very small grains.

Barker \& Cherchneff (1989) and d'Hendecourt et al. (1989) found that
the thermal approximation was valid for computing thermal emission
spectra.
Allamandola, Tielens, \& Barker (1989),
Schutte, Tielens, \& Allamandola (1993),
and Cook \& Saykally (1998)
compared calculations of infrared emission from
PAHs using the thermal
approximation with the results of a more detailed statistical
treatment; they
found that the thermal approximation provided
a good approximation for the emission from low-frequency modes, although
it
overestimated the emission from high frequency modes such as the
$3.3\micron$ C-H stretching mode.
All of these studies used the ``continuous cooling'' approximation
to discuss the overall emission from a PAH following single-photon
heating.

The validity of the ``continuous cooling'' approximation has been
questioned by Siebenmorgen et al. (1992), Manske \& Henning (1998),
and Duley \& Poole (1998).  
Siebenmorgen et al.\ (1992) and Manske \& Henning
(1998) attempted to treat the cooling process as discrete events,
but did not correctly evaluate the transition probabilities, as we show
below in \S\ref{sec:thermal-discrete}.

The present paper addresses the vibrational
excitation of interstellar grains, both small and large, and the
resulting infrared emission.
If energy levels and transition probabilities were known, we could
(at least in principle)
solve for the statistical steady-state populations of the different
energy levels of grains illuminated by a known radiation field.
However, this level of detailed information is generally unavailable,
for even the smallest and simplest PAH molecules.
We therefore construct a ``model PAH'' grain, C$_n$H$_m$,
with a realistic spectrum of energy levels.
Following previous workers (e.g., Schutte et al.\ 1993,
Boulanger et al.\ 1998)
we adopt realistic absorption
cross section $C_\abs(\lambda)$, depending on the number of C atoms,
the H/C ratio, and whether the PAH is neutral or ionized.
We construct a similar model for silicate grains,
with appropriate mode spectrum and absorption cross section.
We divide the energy level spectrum into a manageable number of ``bins'',
estimate the ``bin-to-bin'' transition probabilities,
and can then solve for the steady-state probability $P_j$ of
occupying energy bin $j$.

Our analysis allows for discrete quantum states at low vibrational energies,
and what is effectively a continuum of vibrational levels at large energies.
As we show, the only simplifying assumptions that need to be made are:
(1)
{\it the absorption cross section $C_\abs(\lambda)$ is independent
of the degree of vibrational excitation},
and (2)
{\it internal vibrational relaxation rapidly redistributes internal energy
among the vibrational degrees of freedom of the grain.}
While not exact, these are both expected to be good approximations.
Once $C_\abs(\lambda)$ is adopted 
and the spectrum of vibrational ``normal modes'' (and from this the
vibrational density of states)
is specified, the physics of absorption and emission of radiation
by the grain is essentially fully determined.
With these assumptions, we can solve for the energy distribution function
$P(E)$ for a given grain in an arbitrary radiation field, with full
inclusion of ``multiphoton heating'' effects which become important 
for large grains or intense radiation fields.
Because we solve for $P(E)$ including all radiative transitions,
we obtain the ``exact'' solution against which we can then test solutions
obtained using the
thermal approximation and the continous cooling approximation.

In \S\ref{sec:vibrational_mode_spectrum} we estimate the spectrum of
vibrational modes for small PAH and silicate grains, and obtain the
vibrational density of states for such grains.
The transition probabilities depend on the photon absorption cross sections,
and in \S\ref{sec:C_abs} we describe the cross sections adopted for
PAH grains.

We formulate the ``exact'' statistical treatment in 
\S\ref{sec:statistical_treatment}.
In \S\ref{sec:E(T)} we discuss how the ``temperature'' of a grain is
related to its vibrational energy content,
and in \S\ref{sec:thermal_treatment} we show how the ``thermal''
approximation is related to the exact treatment.
The radiative cooling time for vibrationally-excited grains 
is estimated as a function of $E$ and
grain size in \S\ref{sec:cooling_time}.

In \S\ref{sec:solution_method} we discuss the method used to solve for
the steady-state energy distribution.
In \S\ref{sec:IR_emission} we
show how the infrared emission spectrum can be calculated from the energy
distribution, both for the exact statistical treatment and using the
thermal approximation.
We show some sample solutions
in \S\ref{sec:results}, where we assess the accuracy of the ``thermal''
approximation.
We find the thermal approximation to be quite accurate for calculation
of the overall emission spectrum from a population of very small grains.
Furthermore, we find that the ``continuous cooling'' approximation is
itself quite accurate, and can therefore be used to greatly reduce the amount
of computation required to find the temperature distribution functions.

The above studies therefore provide a methodology for computing the
infrared emission from dust grains irradiated by starlight.
The method is accurate, and can be used for any specified mixture of
grain sizes and compositions; we present results for both carbonaceous
and silicate materials.
We discuss our results in \S\ref{sec:discussion} and summarize the
main points in \S\ref{sec:summary}.

The paper is largely concerned with the physics of infrared emission
from stochastically-heated grains, and practical methods for calculating
the emission spectrum.  Readers who are interested only in the
resulting emission spectrum may wish to proceed directly to 
\S\ref{sec:results}.
In a following paper (Li \& Draine 2001a) we apply this methodology to
construct models to account for the observed infrared emission from
the diffuse interstellar medium.

\section{Vibrational Mode Spectrum and Level Degeneracies
	\label{sec:vibrational_mode_spectrum}}

A grain containing $N_a$ atoms will have 3 translational degrees of
freedom, 3 rotational degrees of freedom, and
$N_m=3N_a-6$ vibrational modes.
Each vibrational mode $j=1,...,N_m$ has a
fundamental frequency $\omega_j$.  
If we approximate the vibrational
modes as harmonic oscillators, then the vibrational energy of the
grain (not including zero-point energy) is
\beq
E=\sum_{j=1}^{N_m} v_j\hbar\omega_j ~~~,
\eeq
where the integer $v_j\geq 0$
is the vibrational quantum number for mode $j$.
If the vibrational system is in thermal equilibrium at temperature $T$,
the expectation value of the energy $E$ is
\beq
\bar{E}(T) = 
\sum_{j=1}^{N_m}\frac{\hbar\omega_j}{\exp(\hbar\omega_j/kT)-1} ~~~.
\label{eq:Esum}
\eeq
where $k$ is Boltzmann's constant.  The heat capacity
\beq
C(T) \equiv \frac{d\bar{E}}{dT} = 
k\sum_{j=1}^{N_m} e^{-\hbar\omega_j/kT} 
\left[\frac{\hbar\omega_j/kT}{1-\exp(-\hbar\omega_j/kT)}\right]^2
~~~.
\label{eq:spec_heat}
\eeq
Ideally, we would have accurate knowledge of the 
$N_m$ frequencies $\omega_j$.
Since the complete 
mode spectrum is known only for a small number of grain candidates
such as C$_{24}$H$_{12}$ coronene (see below), 
we need to estimate the normal mode spectrum
for candidate grains of different sizes and materials.

\subsection{Normal Modes: Polycyclic Aromatic Hydrocarbons
	\label{sec:PAH_normal_modes}}

A polycyclic aromatic hydrocarbon molecule containing $N_\C$ C atoms and
$N_\rmH$ H atoms has $N_m=3(N_\rmH+N_\C-2)$ distinct vibrational modes.
These can be separated into $3(N_\C-2)$ modes
of the C-C skeleton, and $3N_\rmH$ modes associated with the C-H bonds
at the periphery.
The frequencies of these normal modes have been computed for a small
number of polycyclic aromatic hydrocarbons, with some frequencies
determined experimentally, but mode spectra are not yet available for
most PAHs of interest.  Here we 
estimate the mode spectrum for ``generic'' PAHs.
We will treat 5 different types of vibration separately:
out-of-plane C-C modes, in-plane C-C modes, out-of-plane C-H bending,
in-plane C-H bending, and C-H stretching.

Graphite is a useful guide to the C-C modes.
Krumhansl \& Brooks (1953) found that the lattice vibration specific
heat of graphite could be approximated by a model where the
out-of-plane and in-plane C-C vibrations each had a mode spectrum
given by a two-dimensional Debye model, with Debye temperatures
$\Theta_{op}\approx950\K$ for the out-of-plane modes, and 
$\Theta_{ip}\approx2500\K$ for the in-plane modes.
This suggests that the C-C vibrational modes for a PAH might be
similarly approximated.

A $n$-dimensional Debye spectrum has modes uniformly distributed
in $E^n$ up to a maximum energy $k\Theta$.
Consider a mode spectrum of the form
\beq
\hbar\omega_j = k\Theta_D
\left[
\frac{(1-\beta_n)}{N_m}(j-\delta_j) + \beta_n
\right]^{1/n}
~~~j=1,...,N_m.
\label{eq:model_mode_spec}
\eeq
For the PAH C-C modes we set the dimensionality $n=2$, with 
$N_m=N_\C-2$ for the out-of-plane modes, and
$N_m=2(N_\C-2)$ for the in-plane modes.

If $\delta_j$ is independent of $j$, the modes are uniformly distributed
in $E^2$.  We will in fact take
\begin{eqnarray}
\delta_j &=& 1/2 ~~{\rm for}~~j\neq 2 ~{\rm or}~ 3
	\label{eq:delta23}\\
	&=& 1 ~~{\rm for}~~j=2 ~{\rm or}~ 3
	\label{eq:deltaj}
\end{eqnarray}
because this shift of modes 2 and 3 to 
lower frequency improves the match to the actual
mode spectrum of coronene.

As discussed in Appendix \ref{app:geom} we assume PAHs to be planar for
$N_\C\leq 54$, 
approximately spherical for $N_\C > 102$, with intermediate shape for
$54 < N_\C < 102$.
Thus for $N_\C\leq54$ we expect the lowest frequency mode
$\omega_1\propto N_\C^{-1/2}$,
approximately constant $\omega_1$ for $54<N_\C<102$,
and 
$\omega_1\propto N_\C^{-1/3}$ for $N_\C>102$.
We achieve this by taking
\begin{eqnarray}
\beta_2 &=& 0 ~~~{\rm for}~N_\C\leq 54
\nonumber\\
&=& \frac{1}{(2N_m-1)}\left(\frac{N_\C-54}{52}\right) 
~~~{\rm for}~54 < N_\C \leq 102
\nonumber\\
&=& \frac{1}{(2N_m-1)}
\left[\frac{(N_\C-2)}{52}\left(\frac{102}{N_\C}\right)^{2/3}-1\right] 
~~~{\rm for}~N_\C > 102
\label{eq:beta_graph}
\end{eqnarray}
We use equations (\ref{eq:model_mode_spec}-\ref{eq:beta_graph}) 
with $N_m=N_\C-2$ and
$k\Theta_{op}/hc=600\cm^{-1}$ ($\Theta_{op}=863\K$) 
for the out-of-plane C-C modes,
and $N_m=2(N_\C-2)$ and $k\Theta_{ip}/hc=1740\cm^{-1}$
($\Theta_{ip}=2504\K$) 
for the in-plane C-C modes.

We can adequately approximate the C-H stretching and bending modes by
$N_\rmH$ out-of-plane bending modes at 
$\lambda_{\C\rmH,op}^{-1}=(11.3\micron)^{-1}=886\cm^{-1}$,
$N_\rmH$ in-plane bending modes at 
$\lambda_{\C\rmH,ip}^{-1}=(8.6\micron)^{-1}=1161\cm^{-1}$,
and $N_\rmH$ C-H stretching modes at 
$\lambda_{\C\rmH,str}^{-1}=(3.3\micron)^{-1}=3030\cm^{-1}$.

For our model ``astronomical PAH'' sequence, we will assume
\begin{eqnarray}
N_\rmH &=& \INT\left(0.5 N_\C+0.5\right) ~~~~N_\C\leq 25
	\nonumber\\
	&=& \INT\left(2.5 \sqrt{N_\C}+0.5\right) ~~~~25 \leq N_\C \leq 100
	\nonumber\\
	&=& \INT\left(0.25 N_\C+0.5\right) ~~~~~N_\C \geq 100
	\label{eq:H_over_C}
\end{eqnarray}
where $\INT(x)$ is the integer part of $x$.
The pericondensed species coronene C$_{24}$H$_{12}$,
circumcoronene C$_{54}$H$_{18}$, and 
dicircumcoronene 
C$_{96}$H$_{24}$
are members of the sequence prescribed by
eq.\ (\ref{eq:H_over_C}).
Beyond $N_\C=100$ we assume H/C $\approx 0.25$.
For a carbonaceous grain containing $N_\C$ carbon atoms, we associate a
representative ``radius'' $a = 10\Angstrom (N_\C/468)^{1/3}$, corresponding
to the carbon density $\rho=2.24\g\cm^{-3}$ of graphite.

\begin{figure*}[ht]
\centerline{\epsfig{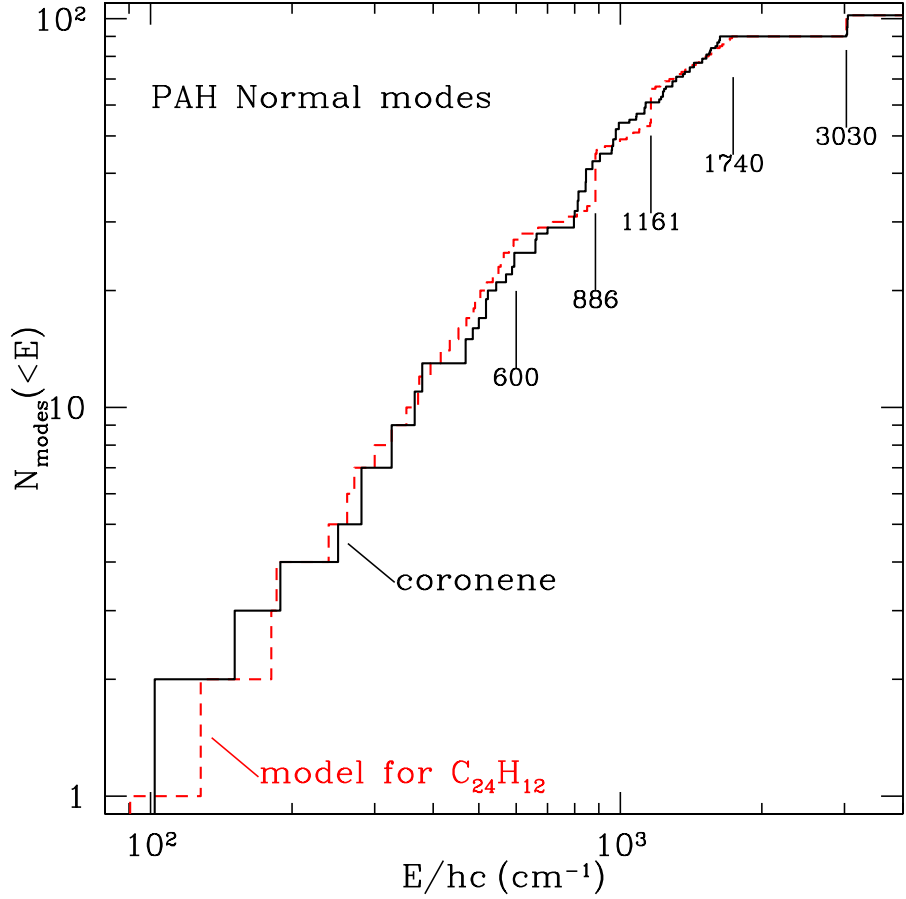}}
\figcaption{\footnotesize
	\label{fig:PAH_mode_spec}
	$N_{\rm modes}(E)$, the number of vibrational modes with energy
	$<E$, for coronene C$_{24}$H$_{12}$ (solid lines)
	compared to the model spectrum (broken lines) 
	given by eq.(\ref{eq:model_mode_spec}-\ref{eq:beta_graph})
	with $N_m=22$ modes with $k\Theta_{op}/hc=600\cm^{-1}$,
	$N_m=44$ modes with $k\Theta_{ip}/hc=1740\cm^{-1}$,
	$N_m=12$ modes with $\lambda^{-1}=886$,
	$1161$, and $3030\cm^{-1}$ (see text).
	}
\end{figure*}

We use C$_{24}$H$_{12}$ coronene, for which the mode spectrum is known
(Cyvin 1982; Cyvin et al. 1984), to test the above model.
As shown in Figure \ref{fig:PAH_mode_spec}, 
our ``synthetic'' mode spectrum is in excellent agreement with the 
actual mode spectrum
of coronene.

\begin{figure*}[ht]
\centerline{\epsfig{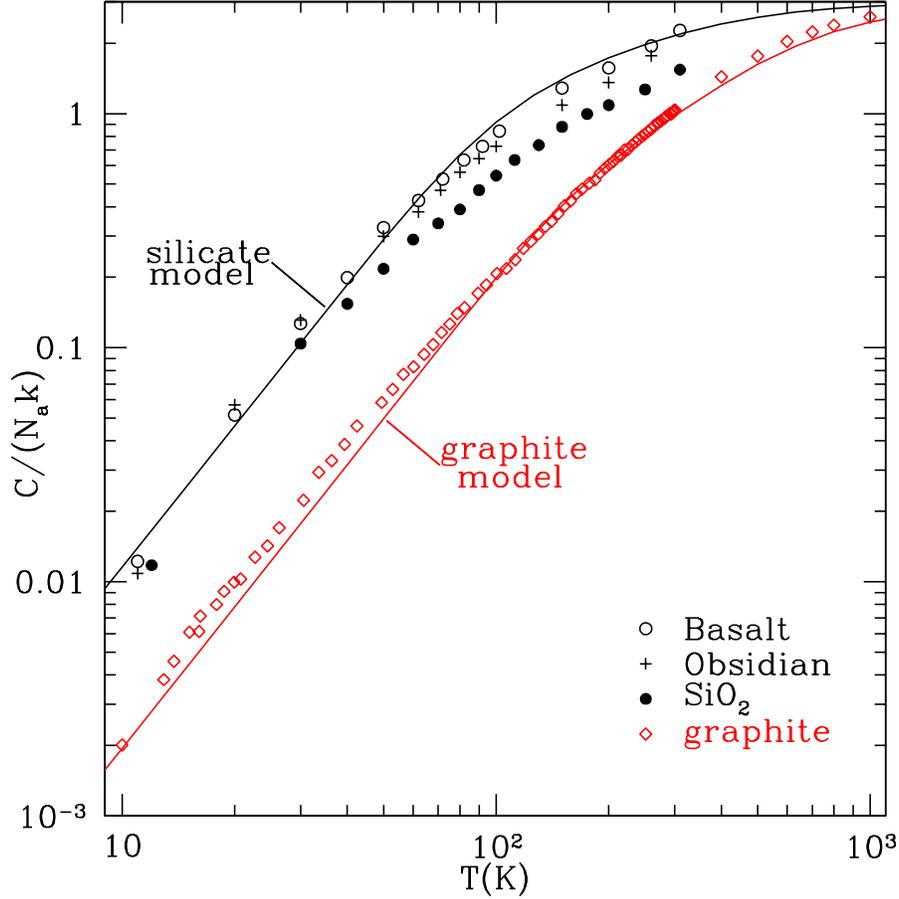}}
\figcaption{\footnotesize
	\label{fig:spec_heats}
	Heat capacity of graphite 
	(deSorbo \& Tyler 1953; Furukawa \& Douglas 1972)
	and model specific heat given by eq.\,(\ref{eq:graph_spec_heat});
	and
	heat capacity of SiO$_2$ glass, obsidian glass,
	and basalt glass (L\'eger, Jura, \& Omont 1985), together with
	the model specific heat given by eq.\,(\ref{eq:sil_spec_heat}).
	}
\end{figure*}

Applying our synthetic mode spectrum for PAHs in the limit
$N_\C\rightarrow\infty$ and $N_\rmH=0$, the mode spectrum of the C-C
modes reduces to the 2-dimensional Debye model.
The heat capacity becomes
\beq
C_{\rm graph}= (N_\C-2) k
\left[
f_2^\prime(T/863\K) + 2 f_2^\prime(T/2504\K)
\right]
\label{eq:graph_spec_heat}
\eeq
\beq
f_n(x) \equiv \frac{1}{n}\int_0^1 \frac{y^n dy}{\exp(y/x)-1} ~~~,~~~
f_n^\prime(x)\equiv \frac{d}{dx}f_n(x) ~~~.
\label{eq:f_n}
\eeq
In Figure \ref{fig:spec_heats} we compare eq.\ (\ref{eq:graph_spec_heat})
with experimental data for (bulk) graphite.
The excellent agreement confirms that our model spectrum for the C-C
vibrational modes applies from small PAH molecules up through
macroscopic samples of graphite.

\subsection{Normal Modes: Silicate Grains
	\label{sec:sil_normal_modes}}

\begin{figure*}[ht]
\centerline{\epsfig{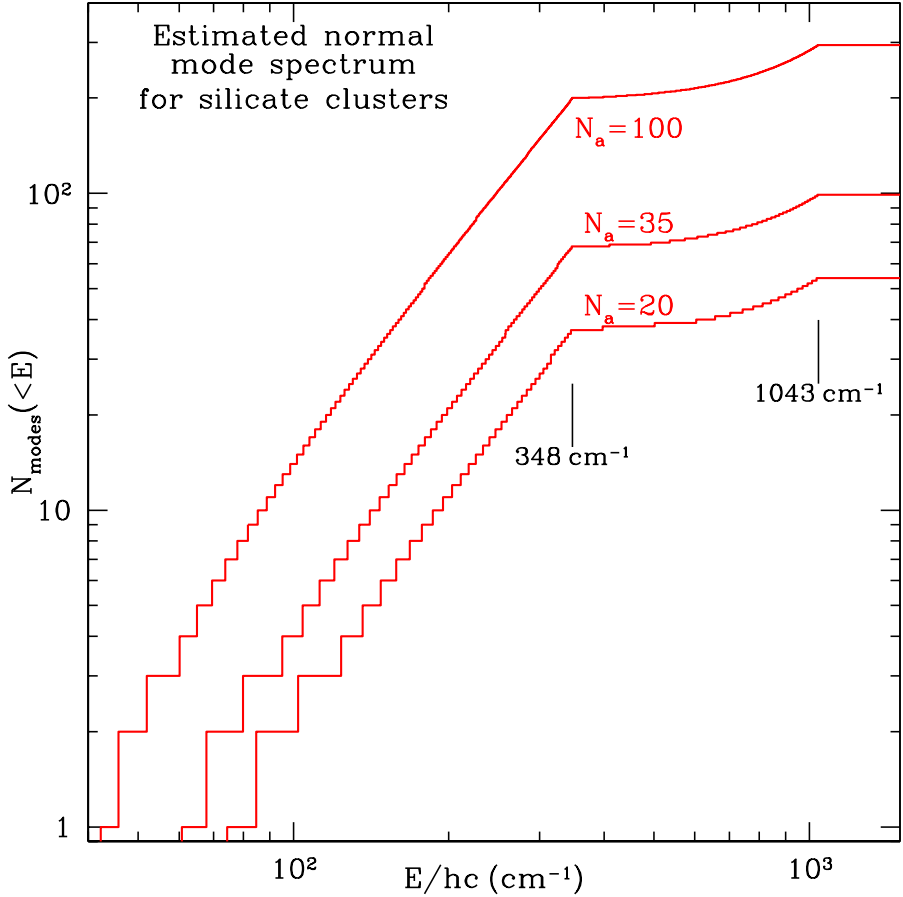}}
\figcaption{\footnotesize
	\label{fig:sil_modes}
	Vibrational mode spectrum adopted for silicate clusters with
	$N_a=20$, 35, and 100 atoms.
	}
\end{figure*}

In the case of silicate grains, we use experimental specific heats
as a guide to the vibrational mode spectrum.
The specifics heat of SiO$_2$ glass, 
obsidian glass (75\% SiO$_2$ and 25\% metal oxides by mass), and
basalt glass (50\% SiO$_2$, 50\% metal oxides by mass) reported
by L\'eger, Jura, \& Omont (1985) are plotted Figure \ref{fig:spec_heats}.
As seen in Figure \ref{fig:spec_heats}, 
the experimental results for bulk silicates (basalt and obsidian)
can be satisfactorily reproduced if 2/3 of the
vibrational modes are distributed according to a Debye model with
$n=2$ and
Debye temperature $\Theta=500\K$, and 1/3 of the modes are described by a
Debye model with $n=3$ and  $\Theta=1500\K$:
\beq
C_{\rm sil} = (N_a-2)k\left[2f_2^\prime(T/500\K)+
f_3^\prime(T/1500\K)\right]
\label{eq:sil_spec_heat}
\eeq
where $N_a$ is the number of atoms in the cluster, and
$f_n$ is given by eq.\ (\ref{eq:f_n}).
Eq.\, (\ref{eq:sil_spec_heat})
agrees quite well with the measured specific heats for
obsidian and basalt, so we use this model to 
estimate the mode spectrum for silicate grains.

We expect the lowest frequency mode $\omega_1 \propto N_a^{-1/3}$.
To achieve this scaling for modes distributed according to an
$n-$dimensional Debye model [eq.\ (\ref{eq:model_mode_spec})], we set
\beq
\beta_n = \frac{N_m^{1-n/3}-1}{2N_m-1} ~~~;
\label{eq:beta_sil}
\eeq
note that for $n=3$ this gives $\beta_3=0$.

For small silicate grains,
two-thirds of the vibrational modes are then approximated by 
eqs.\,(\ref{eq:model_mode_spec}-\ref{eq:deltaj})
with $n=2$, $\Theta_D=500\K$, and $N_m=2(N_a-2)$, and the 
remaining third by
eqs.\,(\ref{eq:model_mode_spec}-\ref{eq:deltaj})
with $n=3$, $\Theta_D=1500\K$, and $N_m=N_a-2$.
The resulting normal mode spectrum is shown in Figure \ref{fig:sil_modes}.

\subsection{Density of States\label{sec:density_of_states}}

\begin{figure*}[ht]
\centerline{\epsfig{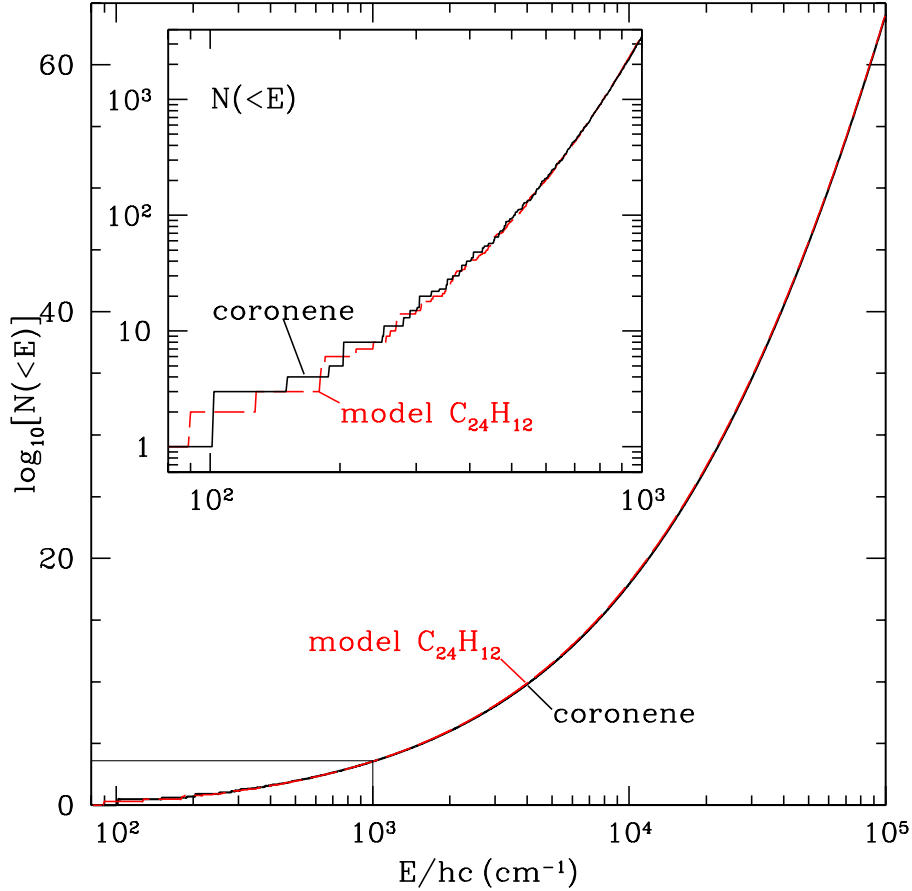}}
\figcaption{\footnotesize
	\label{fig:dens_states}
	Total number of vibrational states $N(E)$ with energy less than $E$
	for coronene, and for the model PAH with $N_\C=24$ and $N_\rmH=12$.
	}
\end{figure*}

Let there be a total of $N_m=3(N_a-2)$ distinct normal modes.
The general vibrational state can then be specified by the
$N_m$-tuple $(v_1,v_2,...,v_{N_m})$ giving the vibrational  
quantum numbers for each of these modes.
If each mode is approximated by a harmonic oscillator, then
$N(E)$, 
the number of distinct vibrational states with total vibrational
energy less than or equal to $E$, can be calculated
using the Beyer-Swinehart algorithm (Beyer \& Swinehart 1973;
Stein \& Rabinovitch 1973).
In Figure \ref{fig:dens_states} we show $N(E)$ computed for C$_{24}$H$_{12}$,
using both the actual normal mode spectrum for coronene and our model normal
mode spectrum for C$_{24}$H$_{12}$ (see Figure \ref{fig:PAH_mode_spec}).
The resulting $N(E)$ are essentially identical for $E/hc\gtsim 300\cm^{-1}$.

\begin{figure*}[ht]
\centerline{\epsfig{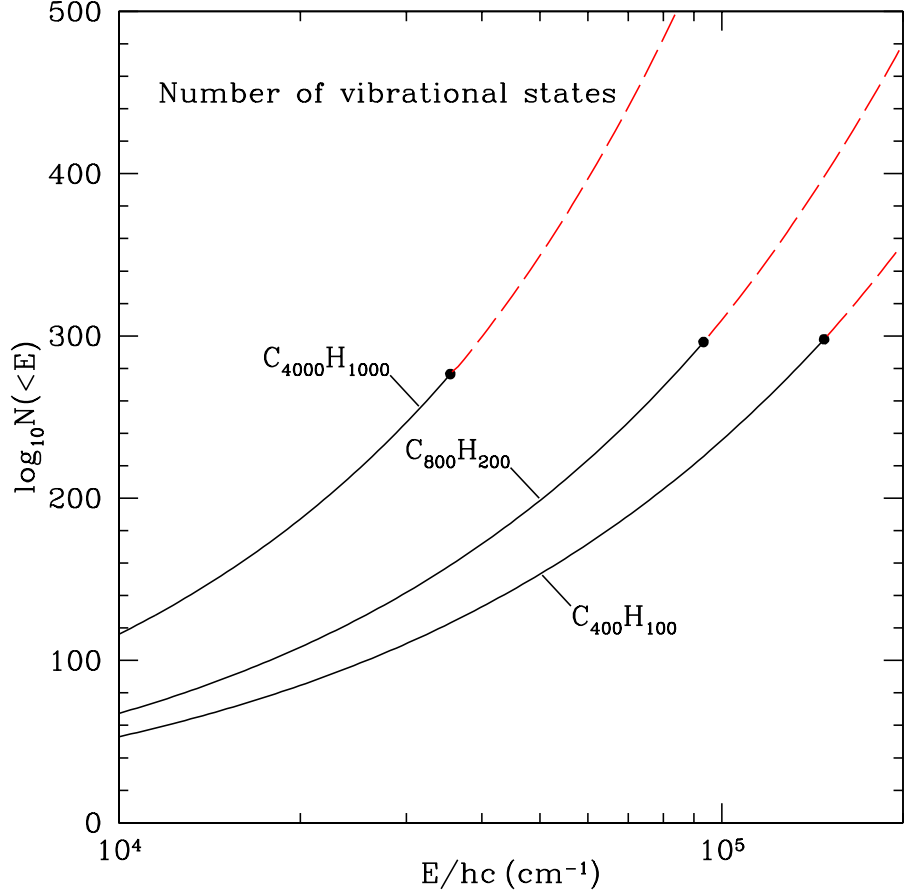}}
\figcaption{\footnotesize
	\label{fig:N(E)extrap}
	Total number of vibrational states $N(E)$ with energy less than
	$E$ for C$_{400}$H$_{100}$ and
	C$_{800}$H$_{200}$.  Solid line is the direct
	Beyer-Swinehart calculation for $N<10^{300}$; broken line is from 
	eq.\,(\ref{eq:int_dE/T}).
	}
\end{figure*}

The Beyer-Swinehart algorithm is remarkably efficient, and with IEEE standard
64-bit
arithmetic can be applied to calculate $N(E)$ up to $E/hc=10^5\cm^{-1}$
for clusters with up to $\sim$600 atoms.
For a given cluster, 
we employ the Beyer-Swinehart algorithm for energies up to
an energy $E_t$ where $N(E_t)\approx 10^{300}$,
beyond which we are limited
by inability to calculate floating point numbers exceeding 
$2^{1024}\approx10^{308}$.
For $E>E_t$ we calculate the density of
states by integrating $dS=dQ/T$:
\beq
\ln N(E) = \ln N(E_t) + \int_{E_t}^E \frac{dE^\prime}{kT(E^\prime)}
\label{eq:int_dE/T}
~~~,
\eeq
where $T(E)$ is the temperature at which the grain has energy $E$ 
(see \S\ref{sec:E(T)}).
Figures \ref{fig:dens_states} and \ref{fig:N(E)extrap} show $N(E)$ evaluated 
for C$_{24}$H$_{12}$, C$_{400}$H$_{100}$, C$_{800}$H$_{200}$, and
C$_{4000}$H$_{1000}$ using our
model density of states.

\section{Absorption Cross Sections
	\label{sec:C_abs}}

\begin{figure*}[ht]
\centerline{\epsfig{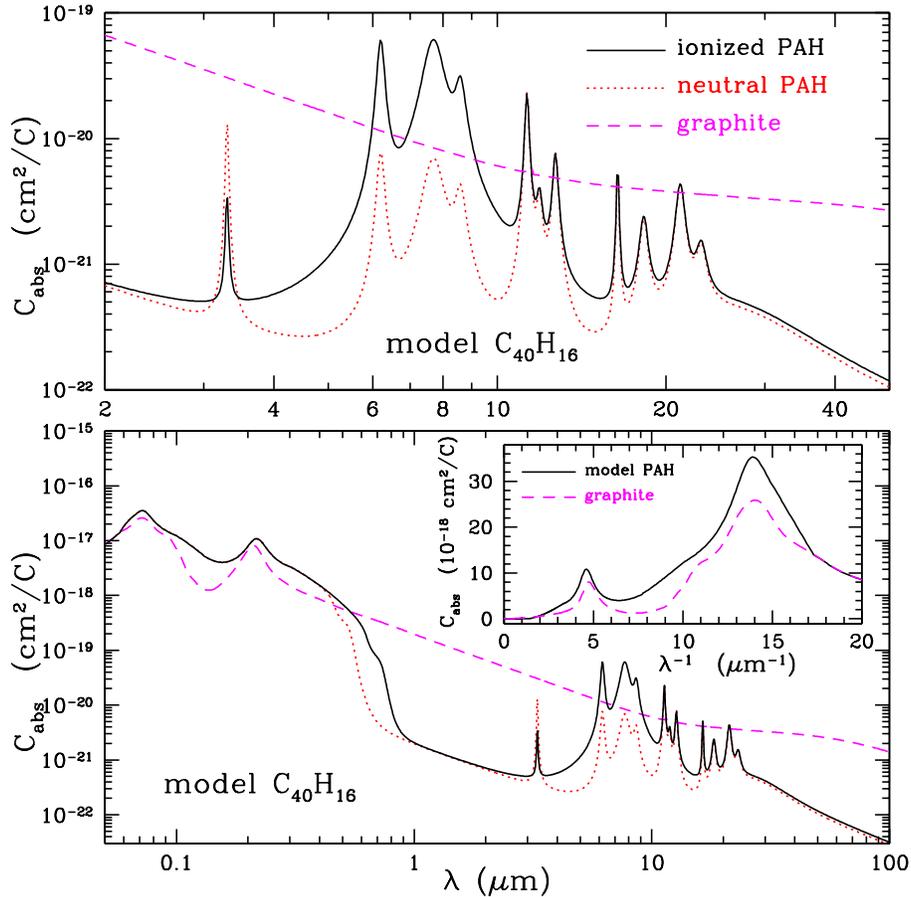}}
\figcaption{\footnotesize
        \label{fig:PAHcs}
        Optical properties of neutral and ionized PAHs,
	using C$_{40}$H$_{16}$ as an example
        (see Li \& Draine 2001a for further information). 
        Also plotted for comparison 
	is the absorption cross section of graphite.
	Note the difference between neutral and ionized PAHs,
	in particular in the 3--12$\micron$ region.
        In the visible to far UV, the principal difference 
        is the location of the visual absorption edge (which is also 
        size-dependent) for ionized PAHs. 
	Insert shows the UV absorption for graphite and PAHs. 
        }
\end{figure*}

Following previous workers 
(e.g., Schutte et al.\ 1993),
we estimate the absorption cross sections for both neutral and
ionized PAHs from far ultraviolet to far infrared 
(see Li \& Draine 2001a for details)
based on available experimental 
data (Allamandola et al.\ 1999; Hudgins \& Allamandola 1999; and references
therein)
and guided by astronomical observations (e.g., Boulanger et al.\ 1998).
In the ultraviolet and infrared the 
resulting cross sections are mainly a collection of 
Drude profiles: the $\sigma-\sigma^{\star}$ transition mode 
($\lambda^{-1}\simeq 14\mu$m$^{-1}$); the $\pi-\pi^{\star}$ transition 
($\lambda^{-1}\simeq 4.6\mu$m$^{-1}$); 
the C-H stretching mode ($\lambda = 3.3\mu$m);
the C-C stretching modes ($\lambda = 6.2, 7.7\mu$m);
the C-H in-plane bending mode ($\lambda = 8.6\mu$m);
the C-H out-of-plane bending modes ($\lambda = 11.3, 11.9, 12.7\mu$m);
and a few weak features attributed to the C-C bending modes 
($\lambda = 16.4, 18.3, 21.2, 23.1\mu$m). In addition to these features,
we have included continuous absorption in the
far-UV, near-UV/visible, as well as a small amount of continuous
absorption in the infrared.
[For further information see Li \& Draine (2001)].
In Figure \ref{fig:PAHcs} we display the adopted absorption cross sections
for neutral and 
ionized PAHs.

Following Li \& Draine (2001), 
we make a smooth transition from PAH optical properties for 
$N_\C\ltsim 6\times10^4$ ($a < 50\Angstrom$)
to graphite properties for $N_\C\gtsim 5\times10^5$ ($a > 100\Angstrom$),
calculated using
Mie theory with the 1/3-2/3 approximation (Draine \& Malhotra 1993) and
dielectric functions from Draine \& Lee (1984).

For silicates, we use Mie theory with dielectric constants for
astronomical silicate (Draine \& Lee 1984).\footnote{
	We use the ``smoothed UV'' silicate dielectric function
	discussed by Weingartner \& Draine 2001a, with modifications to
	the far-infrared emissivity proposed by Li \& Draine 2001a,
	available at
	http://www.astro.princeton.edu/$\sim$draine .
	}

\section{Excitation of Very Small Grains: Exact Statistical Treatment
	\label{sec:statistical_treatment}}

\subsection{Assumptions
	\label{sec:assumptions}}

We characterize the state of the grain by its vibrational energy $E$.
There are too many energy levels to consider individually, so
we group them into $\Nb+1$ ``bins'' $j=0,...,\Nb$,
where the $j$-th bin is
$[E_{j,\min},E_{j,\max})$, with representative energy 
$E_j\equiv(E_{j,\min}+E_{j,\max})/2$, and
width $\Delta E_j\equiv E_{j,\max}-E_{j,\min}$.
The ``degeneracy'' $g_j$ is the number of distinct quantum states
included in bin $j$.
The procedure used for specifying the bins is described in
Appendix \ref{app:energy_bins}.

Let $P_j$ be the probability of finding the grain in bin $j$.
The probability vector $P_j$ evolves according to
\beq
\frac{d}{dt}P_i = \sum_{j\neq i} \bT_{ij} P_j - \sum_{j\neq i} \bT_{ji}P_i 
~~~i=0,1,...,\Nb,
\label{eq:Pdot}
\eeq
where the transition matrix element $\bT_{fi}$ is
the probability per unit time for
a grain in bin $i$ to make a transition to one of the levels in bin $f$.
All of the physics is embodied in the transition matrix $\bT$.

A truly exact treatment of the grain excitation would require knowledge of
all the state-to-state transition probabilities between energy levels.
Given the large number of energy levels to be considered, this is utterly
infeasible at this epoch.
In order to estimate the required state-to-state rates,
we make only the following three assumptions:
\begin{itemize}
\item The grain absorption cross section $C_{\abs}(h\nu)$
	depends on the photon energy $h\nu$,
	but does {\it not} depend on the vibrational energy 
	of the grain prior to the absorption.
\item The vibrational modes of the grain can be approximated by
	harmonic oscillators.
\item The energy of absorbed photons is distributed ergodically 
	among the $3N_a-6$
	vibrational degrees of freedom before any infrared emission takes
	place.\footnote{
		Since the grain will in general have angular momentum, and
		the principal values of the moment of inertia tensor will
		generally be nondegenerate, in principle some of the heat
		can appear in rotational excitation 
		(Lazarian \& Draine 1999a,b).
		However, since there are only
		three rotational degrees of freedom, this is at most a small
		effect, and will be neglected.}
\end{itemize}
We now show that
with only these three assumptions, together with knowledge of
\begin{itemize}
\item the spectrum of fundamental vibrational modes,
\item the grain absorption section $C_{\abs}(h\nu)$, and
\item the starlight energy density $u_E$,
\end{itemize}
we can fully determine the emission from interstellar dust grains.
Note that the notion of ``grain temperature'' does not appear in 
what we will refer to as the ``exact-statistical treatment''.

\subsection{Upward Transitions\label{sec:upward_transitions}}

In Appendix \ref{app:T_fi} we show that the rate for upward transitions
$l\rightarrow u$ is
\begin{eqnarray}
\bT_{ul} &=& 
\frac{c \Delta E_u}{E_u-E_l}
\int_{W_1}^{W_4} G_{ul}(E) C_{\abs}(E) u_E dE
~~~~~~~{\rm for}~ u<\Nb
\label{eq:T_ul}
\\
\bT_{Ml}&=& 
\frac{c}{E_M-E_l}
\left[
\int_{W_1}^{W_c} \left(\frac{E-W_1}{W_c-W_1}\right) C_{\abs}(E) u_E dE +
\int_{W_c}^\infty 
 C_{\abs}(E) u_E dE
\right]
\label{eq:T_Ml}
\end{eqnarray}
\begin{eqnarray}
G_{ul}(E)&\equiv&\frac{(E-W_1)}{\Delta E_u \Delta E_l} 
		~~~{\rm for}~W_1<E<W_2
		\\
	&\equiv&\frac{\min\left[\Delta E_l,\Delta E_u\right]}
		{\Delta E_u\Delta E_l}
		~~~{\rm for}~W_2<E<W_3
		\\
	&\equiv&\frac{(W_4-E)}{\Delta E_u\Delta E_l} 
		~~~{\rm for}~W_3<E<W_4\label{eq:G_ul}
		\\
	&\equiv&0 ~~~ {\rm for}~ E<W_1 ~{\rm or}~E>W_4
		\\
W_1&\equiv& E_{u,\min}-E_{l,\max}\\
W_2&\equiv& \min\left[(E_{u,\min}-E_{l,\min}),(E_{u,\max}-E_{l,\max})\right]\\
W_3&\equiv& \max\left[(E_{u,\min}-E_{l,\min}),(E_{u,\max}-E_{l,\max})\right]\\
W_4&\equiv& E_{u,\max}-E_{l,\min} \label{eq:W_4}\\
W_c&\equiv& E_{M,\min}-E_{l,\min}
\end{eqnarray}
and $u_E dE$ is the energy density due to photons with energies in
$[E,E+dE]$.
Note that eq.\ (\ref{eq:T_Ml}) takes 
energy absorbed in transitions 
to levels beyond the highest bin and allocates it
to the highest bin ($M$).
In practice, we ensure that the highest bin is set high enough that
the rate of such transitions is negligible, and
the population $P_{\Nb}$ is negligibly small.

The factor $G_{ul}(E)$ represents the correction for finite bin width;
if $\max(\Delta E_l, \Delta E_u) \ll (E_u-E_l)$, then eq.\ (\ref{eq:T_ul})
becomes
\begin{eqnarray}
\bT_{ul}&\approx& cC_{\abs}(E) \frac{u_E \Delta E_u }{E_u-E_l}
~~~{\rm for}~u< \Nb\\
\bT_{Ml}&\approx& cC_{\abs}(E) \frac{u_E\Delta E_M }{E_M-E_l}
+ \frac{c}{E_M-E_l}\int_{E_M-E_l}^\infty 
C_\abs(E) u_E dE
~~~.
\end{eqnarray}
Most previous studies (e.g., Guhathakurta \& Draine 1989) 
have used this approximate form, but
eq.\ (\ref{eq:T_ul}-\ref{eq:W_4}) 
make proper allowance for finite bin width and
are to be preferred.
Correction for finite bin width is important when the treatment is
applied to grains with radii $a\gtsim50\Angstrom$, since for tractable
numbers of bins $M\ltsim10^3$, many bins will necessarily be broad, with
$\Delta E_u \gtsim kT_u$.

For the special case of transitions $u\!-\!1\rightarrow u$ we include
``intrabin'' absorptions:
\beq
\bT_{u,u\!-\!1} = 
\frac{c}{E_u\!-\!E_{u\!-\!1}}
\left[
\int_0^{W_4} \Delta E_u G_{u,u-1}(E) C_{\abs}(E) u_E dE 
+ 
\int_0^{\Delta E_{u\!-\!1}} 
\left(1-\frac{E}{\Delta E_{u\!-\!1}}\right) 
C_{\abs}(E) u_E dE
\right]
\label{eq:Tu,u-1}
\eeq
but the ``intrabin'' contribution (second integral in eq.\ \ref{eq:Tu,u-1})
is negligible in all cases of interest.

\subsection{Downward Transitions}

The radiative coupling between levels is determined by
the absorption cross section $C_\abs(h\nu)$.
The Einstein
$A$ coefficient for radiative decay $u\rightarrow l$, averaged over the
sublevels in bin $u$, is (see Appendix \ref{app:T_fi})
\beq
\bT_{lu} = \frac{8\pi}{h^3c^2} \frac{g_l}{g_u}
\frac{\Delta E_u}{E_u-E_l} 
\int_{W_1}^{W_4} G_{ul}(E) E^3 C_{\abs}(E)
\left[1+\frac{h^3c^3}{8\pi E^3}u_E\right] dE ~~{\rm for}~l<u-1 .
\label{eq:T_lu}
\eeq
The degeneracies 
$g_u$ and $g_l$ are the numbers of energy states in bins $u$ and
$l$, respectively (see \S\ref{sec:density_of_states}):
\beq
g_j\equiv N(E_{j,\max})-N(E_{j,\min}) \approx \left(\frac{dN}{dE}\right)_{E_j}
\Delta E_j
~~~.
\label{eq:g_j}
\eeq
The term proportional to $u_E$ in eq.\ (\ref{eq:T_lu})
is the contribution of stimulated
emission.
For transitions $u\rightarrow u-1$ to the adjacent bin 
we have $W_1=0$: even very long wavelength emission can contribute
if $C_{\abs}>0$.
For this case we must also include the power radiated in ``intrabin''
transitions within bin $u$:
\begin{eqnarray}
\bT_{u-1,u}&=&\frac{8\pi}{h^3c^2}\frac{g_{u-1}}{g_u}
\frac{\Delta E_u}{E_u-E_{u-1}}
\int_0^{W_4}G_{u,u-1}(E)E^3 C_\abs (E)
\left[1+\frac{h^3c^3}{8\pi E^3}u_E\right] dE +
\nonumber
\\
&&
\frac{8\pi}{h^3c^2}\frac{1}{E_u-E_{u-1}}
\int_0^{\Delta E_u} \left(1-\frac{E}{\Delta E_u}\right)
E^3 C_\abs (E) \left[1+\frac{h^3c^3}{8\pi E^3}u_E\right] dE
\label{eq:intrabin_stat}
\end{eqnarray}
Thus we see that we require only $C_{\abs}(E)$, the degeneracies
$g_j$, and the starlight spectrum $u_E$ to completely 
determine the transition matrix $\bT_{fi}$.
We stress that ``grain temperature'' does not enter the exact-statistical
treatment (e.g., eq.\,\ref{eq:T_ul},\ref{eq:T_lu}).
In \S\ref{sec:thermal_treatment} below we will discuss ``thermal'' 
approximations to estimate the downward transition rates $\bT_{lu}$.
We first discuss an estimate $T(E)$ for the ``temperature'' of
a state of specified vibrational energy $E$.

\section{Energy vs. Temperature \label{sec:E(T)}}

\subsection{``Exact'' Mode Spectrum}

Consider a grain with vibrational energy $E_u$.
If the grain is large, we know that the emission can be estimated from
the grain ``temperature''.
We can
use eq.\,(\ref{eq:Esum}) to
define a ``temperature'' $\tilde{T}(E_u)$ such that $\bar{E}(\tilde{T})=E_u$,
where $\bar{E}(T)$ is the expectation value for the vibrational energy 
$E$ when in contact with
a heat bath at temperature $T$.

\begin{figure*}[ht]
\centerline{\epsfig{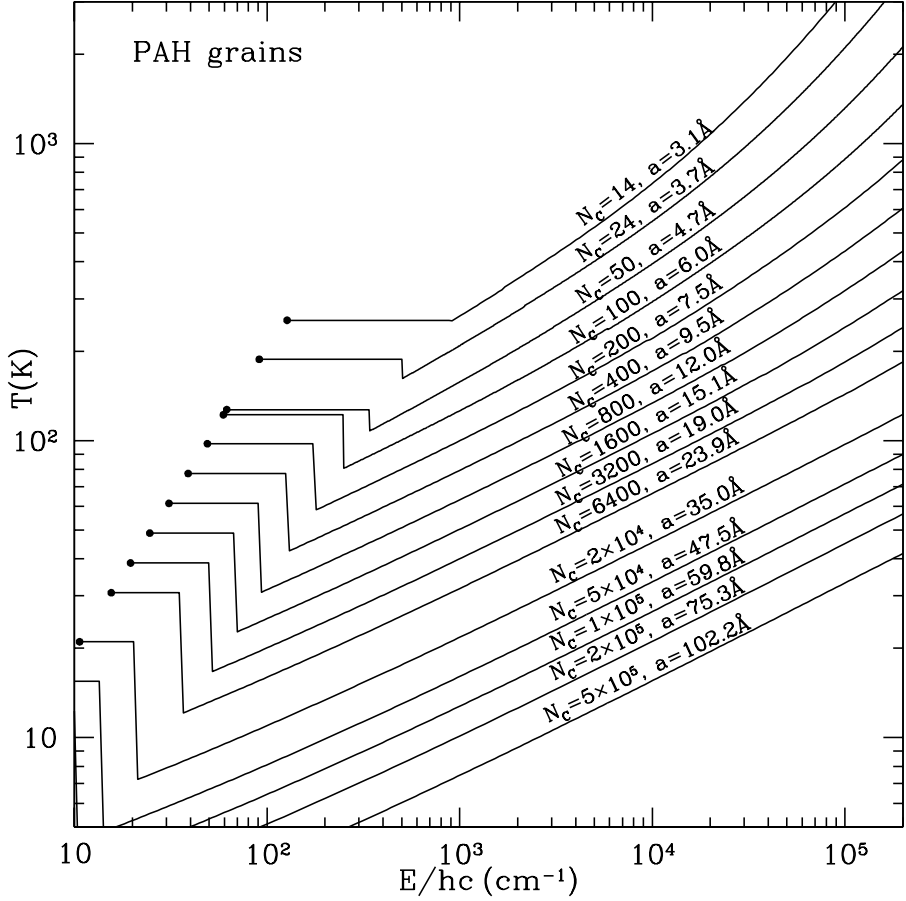}}
\figcaption{\footnotesize
	\label{fig:TvsE_PAH}
	Temperature $T(E)$ for selected PAH grains with
	vibrational energy $E$.
	For each grain size the dot indicates the energy of the
	lowest vibrational mode.
	For each grain, $T(E)$ is discontinuous at $E=\hbar\omega_{20}$
	[see eq.\,(\ref{eq:T_u})].
	}
\end{figure*}

\begin{figure*} [ht]
\centerline{\epsfig{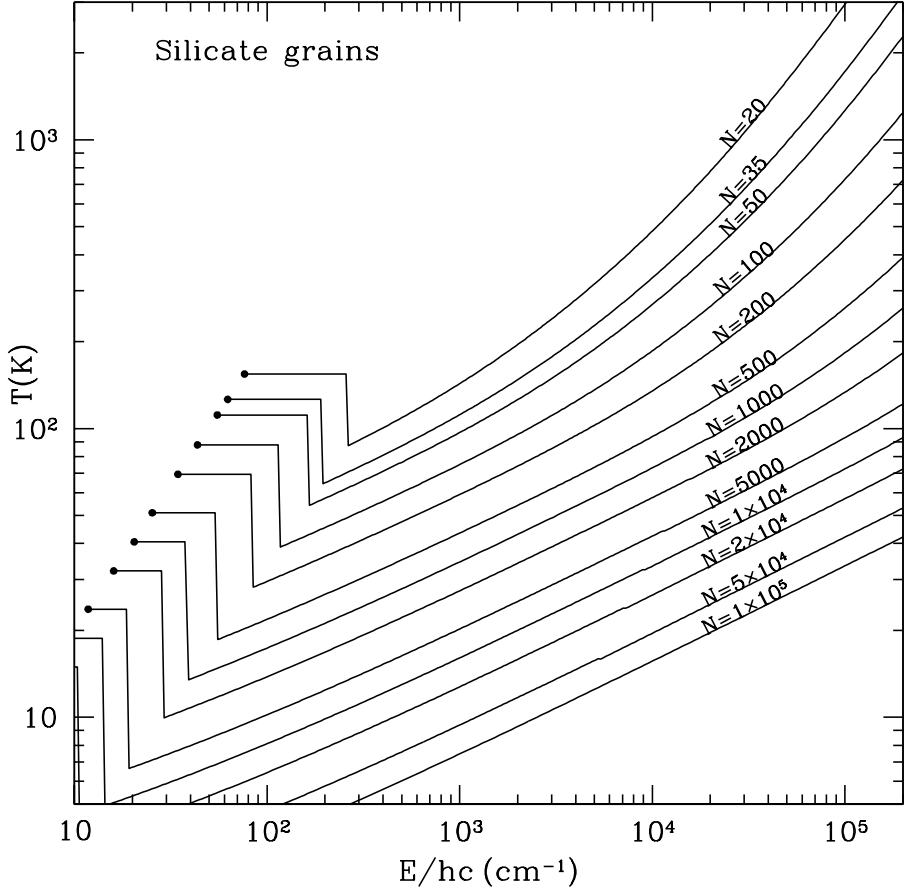}}
\figcaption{\footnotesize
	\label{fig:TvsE_sil}
	As in Fig.\ \ref{fig:TvsE_PAH}, but for silicate grains containing
	$N$ atoms.
	}
\end{figure*}

When there are only a few vibrational quanta in the grain, however,
``temperature'' is not a well-defined concept.
As the most extreme example, 
consider a grain where only the first vibrational mode
is excited, with exactly one quantum: $E_u=\hbar\omega_1$.
If the grain has many vibrational degrees of freedom, then the temperature
$\tilde{T}$ may be quite low (since it is defined in terms of the
expectation value of the vibrational energy summed over all the modes).
However, our grain {\it is} vibrating in the fundamental vibrational mode
and we want to estimate the emission from this mode.
One estimate for the temperature characterizing the excitation of this mode
would be the temperature for which the occupation number {\it of this mode}
is equal to 1.
We will see in \S\ref{sec:thermal_treatment} that when this
``temperature'' is used in the ``thermal approximation'', one obtains the
correct rate of emission in the $1\rightarrow0$ transition.
In fact, we will use this value of the temperature for all energies up to
the energy of the 20th vibrational mode.
For large grains and large energies, however, $\tilde{T}$ is appropriate.
To encompass these two limits, we take the temperature $T_u$
of vibrational level $u$ to be\footnote{
	For $T=\hbar\omega/k\ln2$ a quantized 
	harmonic oscillator has occupation number
	equal to 1.
	}
\begin{eqnarray}
T_u &=& \frac{\hbar\omega_1}{k\ln2} ~~~{\rm for}~ E_u \leq \hbar\omega_{20}
	\nonumber\\
	&=& \tilde{T}(E_u) ~~~{\rm for}~ E_u > \hbar\omega_{20}
	\label{eq:T_u}
\end{eqnarray}
While the choice of $\hbar\omega_{20}$ as the energy separating these
two approximations is arbitrary, we remind the reader that the very notion
of ``temperature'' is questionable for such low degrees of excitation.
Our use of eq.\,(\ref{eq:T_u}) is in fact guided by the fact that the
``thermal approximation'' of \S\ref{sec:thermal_treatment}, when used with
this prescription for $T_u$, gives emission rates in fair agreement
with the ``exact-statistical'' treatment (\ref{eq:T_lu}).
Figures \ref{fig:TvsE_PAH} and
\ref{fig:TvsE_sil} show our adopted 
$T(E)$ for selected PAHs and silicates.\footnote{
	The discontinuity in $T(E)$ at $E=\hbar\omega_{20}$ is obviously 
	unphysical, but we note that the discontinuity is only appreciable
	for large grains at very low energies, and these make a
	negligible contribution to the overall emission spectrum.
	}


\subsection{Debye Mode Spectrum\label{sec:debye_mode_spectrum}}

When the number of modes is large, the summation in equation (\ref{eq:Esum})
contains many terms.
In this case, for PAHs we replace the sums over the C-C modes by 
the (continuum) Debye model discussed above:
\beq
\bar{E}_{\rm PAH}(T) \approx
N_\rmH \sum_{j=1}^3 \frac{\hbar\omega_j}{\exp(\hbar\omega_j/kT)-1}
+
(N_\C -2)\left[ k\Theta_{op} f_2(T/\Theta_{op}) + 
2k\Theta_{ip} f_2(T/\Theta_{ip})\right] ~~~,
\label{eq:E_PAH_Debye}
\eeq
where $f_2$ is defined by eq.\,(\ref{eq:f_n}),
and the summation index $j=1-3$ 
runs over the C-H out-of-plane bending, in-plane
bending, and stretching modes
(see \S\ref{sec:PAH_normal_modes}).
Similarly, for silicate grains we take
\beq
\bar{E}_{\rm sil}(T) \approx
(N_a-2)\left[2k\Theta_2 f_2(T/\Theta_2) + k\Theta_3 f_3(T/\Theta_3)\right]
\label{eq:E_sil_Debye}
\eeq
where $\Theta_2=500\K$ and $\Theta_3=1500\K$.
We use eq.\,(\ref{eq:E_PAH_Debye}) and (\ref{eq:E_sil_Debye}) for
$E>N_a E_1$.

\section{Thermal Approximations for Spontaneous Emission
	\label{sec:thermal_treatment}}

Given the above assumptions about the vibrational mode spectrum 
(\S\ref{sec:vibrational_mode_spectrum}),
the assumption that the absorption cross section depends only
on the photon energy
(\S\ref{sec:C_abs}),
and the assumption of rapid internal vibrational redistribution,
the ``statistical'' approach described above 
(\S\ref{sec:statistical_treatment}) is ``exact''.
It is often desirable, however, to use an alternative ``thermal''
approximation which is less computationally demanding.

The thermal approach has been frequently used
(Greenberg 1976;
Purcell 1976;
Aannestad \& Kenyon 1979;
L\'eger \& Puget 1984; 
Draine \& Anderson 1985; 
Puget et al. 1985;
D\'esert et al. 1986;
Dwek 1986; 
Guhathakurta \& Draine 1989;
Aannestad 1989;
Manske \& Henning 1998).
The only difference between the thermal approach and the statistical
approach concerns the calculation of the downward transition rates
$\bT_{lu}$ which, in contrast to the exact-statistical treatment
[eq.\,(\ref{eq:T_lu})], uses the notion of ``grain temperature''.

We will consider two different approximations which 
make use of the thermal approximation, and compare results for energy
distributions and infrared emission spectra.

\subsection{``Thermal-Discrete'' Approximation
	\label{sec:thermal-discrete}}

In the ``thermal-discrete'' approximation we use 
the same energy bins used in the exact statistical treatment discussed in
\S\ref{sec:statistical_treatment}, and allow the same upward and
downward discrete transitions. However,
instead of using eq.(\ref{eq:T_lu}) to evaluate the downward transition
probabilities ${\bf T}_{lu}$, we replace 
\beq
\frac{g_l}{g_u} \rightarrow 
\frac{\Delta E_l}{\Delta E_u}\frac{1}{\exp(h\nu/kT_u)-1}
\eeq
where $T_u$ is the ``temperature'' of the upper level.
With this replacement, the downward transition rate becomes,
for $l<u-1$:\footnote{
	Siebenmorgen et al.\ (1992) discussed both discrete heating and
	cooling.  However, their downward transition elements were
	evaluated by integrating over the {\it initial}, rather than
	final, enthalpy interval, and hence are in error by a factor
	$\sim \Delta E_u/\Delta E_l$, which is large when
	$\Delta E_u \gg \Delta E_l$, as is often the case.
	Manske \& Henning (1998) also discussed 
	discrete cooling. 
	In the present notation their equation (4)
	becomes
	\begin{eqnarray}
		{\bf T}_{lu} &=& 
		\frac{8\pi}{h^3c^2} \int_{E_u-E_{l,\max}}^{E_u-E_{l,\min}}
               \frac{E^3/\Delta E_u} 
               {\exp\left(E/kT_u\right)-1}
		C_\abs(E)dE ~~~.
	\nonumber
	\end{eqnarray}
	Comparison with eq.\ (\ref{eq:T_lu,thermal}) reveals that
	Manske \& Henning's integrand is off
	by a factor $\sim (E_u-E_l)/[\Delta E_u\Delta E_l G_{ul}(E)]
	\approx (E_u-E_l)/\min(\Delta E_l,\Delta E_u)$.
	This is a large error for the common case
	where the photon energy is large compared to the energy
	bin width, $E\gg\min(\Delta E_l,\Delta E_u)$.
	}
\beq
\bT_{lu}\td  = \frac{8\pi}{h^3c^2} 
\frac{\Delta E_l}{E_u-E_l}
\int_{W_1}^{W_4} G_{ul}(E) 
\frac{E^3 C_{\abs}(E) }{\exp(E/kT_u)-1} 
\left[1+\frac{h^3c^3}{8\pi E^3}u_E\right]
dE
~~~{\rm for}~0<l<u-1 ~.
\label{eq:T_lu,thermal}
\eeq
For $l=u-1$ we include the energy-loss in ``intrabin'' transitions,
as in eq. (\ref{eq:intrabin_stat}):
\begin{eqnarray}
\bT_{u-1,u}\td  &=& \frac{8\pi}{h^3c^2} 
\frac{\Delta E_l}{E_u-E_l}
\int_{0}^{W_4} G_{u,u-1}(E) 
\frac{E^3 C_{\abs}(E)}{\exp(E/kT_u)-1} 
\left[1+\frac{h^3c^3}{8\pi E^3}u_E\right]
dE +
\nonumber\\
&&
\frac{8\pi}{h^3c^2}
\frac{1}{E_u-E_{u-1}}
\int_0^{\Delta E_u} \left(1-\frac{E}{\Delta E_u}\right)
\frac{E^3 C_{\abs}(E)}{\exp(E/kT_u)-1}
\left[1+\frac{h^3c^3}{8\pi E^3}u_E\right]
dE ~.
\label{eq:intrabin_thermal}
\end{eqnarray}

We refer to eq.\ (\ref{eq:T_lu,thermal},\ref{eq:intrabin_thermal})
as the ``thermal-discrete'' approximation.
The advantage of the ``thermal-discrete'' approximation is that one
does not require explicit knowledge of the degeneracies $g_u$ and $g_l$
appearing in equation (\ref{eq:T_lu},\ref{eq:intrabin_stat}).
We obtain the temperature $T_u$ from eq.\,(\ref{eq:T_u}).\footnote{%
	Note that the thermal approximation (\ref{eq:T_lu,thermal})
	reduces to the statistical treatment (\ref{eq:T_lu})
	if the temperature is taken to be
	\begin{eqnarray}
	T_u &=& 
	\frac{h\nu/k}{\ln\left[1+(dN/dE)|_{E_u}/(dN/dE)|_{E_u-h\nu}\right]}
	\nonumber
	\end{eqnarray}
	but this definition of $T_u$ depends on both $u$ and the emitted
	photon energy $h\nu$ -- only in the limit of infinite degrees of
	freedom does $T_u$ no longer depend on the value of $h\nu$.
	}

As discussed in \S\ref{sec:assumptions}, the vibrational ground state
is in a ``bin'' with energy $E_0=0$ and width $\Delta E_0=0$, for which
eq.\ (\ref{eq:T_lu,thermal}) would give $\bT_{0u}\td =0$.
There is really no ``correct'' way to approximate the energy levels
as continuous when dealing with the ground state.
We simply replace $\Delta E_0$ in eq.\ (\ref{eq:T_lu,thermal}) by
$\Delta E_1$.\footnote{%
	If this seems rather arbitrary, we remind the reader that the
	``correct'' approach to the problem is the statistical treatment
	given in \S\ref{sec:statistical_treatment}.
	The ``thermal approximation'' considered here is inappropriate
	when discussing low degrees of excitation.
	}
Thus we do not use eq.\ (\ref{eq:T_lu,thermal}) for transitions
to the ground state, and instead take\footnote{
	It can be verified that if we use $\Delta E_1$ for the width of
	bins 0 and 1, and $g_0=g_1=1$, 
	then $T_1=E_1/k\ln2$ from eq.\,(\ref{eq:T_u})
	gives the same transition rate $\bT_{01}$ as the exact-statistical
	treatment.}
 
\beq
\bT_{0u}\td  = \frac{8\pi}{h^3c^2} 
\frac{\Delta E_1}{\Delta E_u}
\frac{1}{E_u}
\int_{E_{u,\min}}^{E_{u,\max}}
\frac{E^3}{\exp(E/kT_u)-1} 
\left[1+\frac{h^3c^3}{8\pi E^3}u_E\right]
C_{\abs}(E) dE ~~~.
\label{eq:T_0u,thermal}
\eeq

\subsection{``Thermal-Continuous'' Cooling Approximation
	\label{sec:thermal-continuous}}

In all of the above discussion, the grain is assumed to make
discrete transitions to energy levels $l<u$ by emission of single photons --
so-called ``discrete cooling''.
With discrete cooling included, the transition 
matrix $\bT_{fi}$ is generally nonzero
through the upper triangular portion $i>f$.

There are substantial computational advantages 
(see \S\ref{sec:solution_method} below) if
the cooling of the grains is approximated as continuous rather than discrete,
so that the only downward transition from a level $u$ is to the adjacent
level $u-1$.
We refer to this as the ``thermal continuous'' cooling approximation.
We take
\begin{eqnarray}
\bT_{lu}\tc  &=& 0 ~~~{\rm if}~ l < u-1
\\
\bT_{u-1,u}\tc  &=& \frac{1}{(E_u-E_{u-1})}
\sum_{l=0}^{u-1} (E_u-E_l) \bT_{lu}\td
\label{eq:T^cont}
\end{eqnarray}
Using (\ref{eq:T_lu,thermal}), eq.(\ref{eq:sum_del_Gul}), 
and eq.\,(\ref{eq:sum_Eul}),
the transition rate (\ref{eq:T^cont}) from levels $u>2$ 
can be approximated
\beq
\bT_{u-1,u}\tc  \approx \frac{1}{(E_u-E_{u-1})}
\frac{8\pi}{h^3c^2} \int_0^{E_u} \frac{E^3 C_\abs(E)}{\exp(E/kT_u)-1}dE
~~~
{\rm for}~u>1.
\label{eq:T^cont.sum}
\eeq

Because of the computational advantages
(see \S\ref{sec:continuous-cooling-solution}), 
the continuous cooling approximation
has been often used in previous studies (e.g.,
Guhathakurta \& Draine 1989).
While clearly appropriate for
large systems with many degrees of freedom (for which $kT\ll E$ and
emission of one photon does not result in a large reduction in temperature),
validity of the
continuous cooling approximation for
grains with $\ltsim 10^2$ atoms is certainly not
obvious {\it a priori}.
Manske \& Henning (1998) in fact 
argued that the continuous cooling approximation
could lead to significant error.
We will examine the accuracy of the continuous-cooling approximation
in \S\ref{sec:results} below, where we will show that it is in fact
surprisingly accurate even for grains with as few as $\sim30$ atoms.

\section{Cooling Time
	\label{sec:cooling_time}}

\begin{figure*}[ht]
\centerline{\epsfig{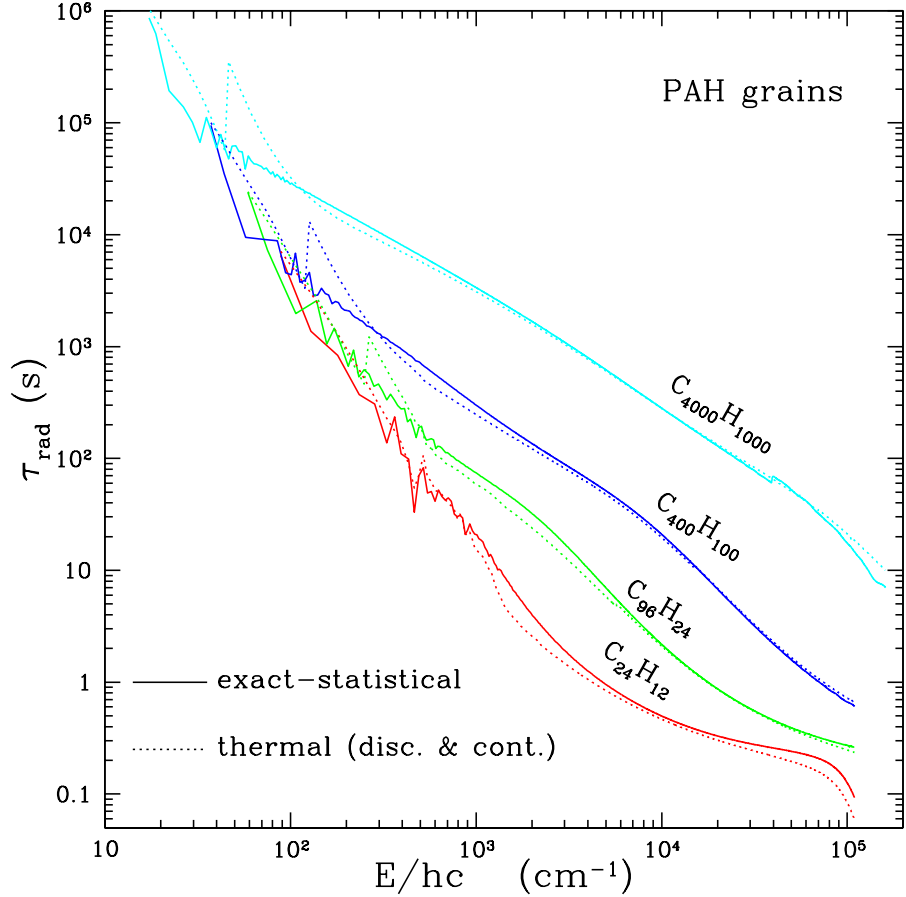}}
\figcaption{\footnotesize
        \label{fig:PAH_cooling_time}
	The radiative cooling time $\tau_\rad$
	as a function of vibrational energy $E$
	for PAH grains, calculated with the
	{\it exact-statistical} and {\it thermal-discrete} cooling models.
	The lowest energy $E$ shown for each grain corresponds to the
	lowest vibrationally-excited state $\hbar\omega_1$.
	The discontinuity in $\tau_\rad$ for the thermal-discrete
	model is due to the different definition of the grain
	``vibrational temperature'' for the lowest 20 excited levels
	(see \S\ref{sec:E(T)} and Fig.\ \ref{fig:TvsE_PAH}).
 	Note that $\tau_\rad$ for the {\it thermal-continuous} model 
        is identical to that for the {\it thermal-discrete} model
        [eq.\,(\ref{eq:T_ul}) and 
	eq.\,(\ref{eq:T_lu,thermal})].
        }
\end{figure*}

\begin{figure*}[ht]
\centerline{
	\epsfig{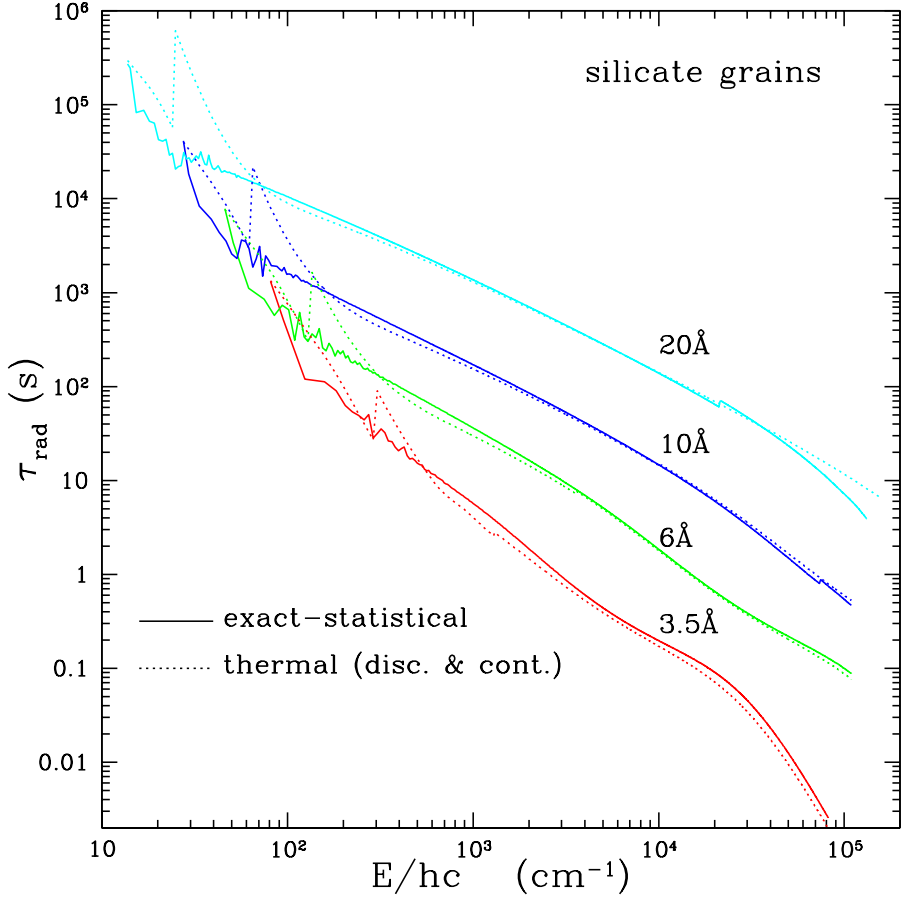}
	}
\figcaption{\footnotesize
        \label{fig:sil_cooling_time}
	Same as Fig.\,\ref{fig:PAH_cooling_time} but for silicate grains.
        }
\end{figure*}

The absorption cross section $C_\abs(E)$ determines both the
rate of heating by starlight and the rate of cooling by infrared emission.
It is useful to define a ``radiative cooling time''
\begin{eqnarray}
\tau_\rad(E_u) &\equiv& \frac{E_u}{\sum_{l<u}\bT_{lu}(E_u-E_l)}
\\
&\approx& \left[
\frac{8\pi}{h^3c^2}\frac{1}{E_u}
\frac{1}{(dN/dE)_u}
\int_0^{E_u} x^3 C_\abs(x) \left[1+\frac{h^3c^3}{8\pi x^3}u_x\right]
\left(\frac{dN}{dE}\right)_{E_u-x} dx
\right]^{-1}
\label{eq:tau_rad_approx}
\end{eqnarray}
where eq.\,(\ref{eq:sum_Eul}) has been used to obtain
eq.\,(\ref{eq:tau_rad_approx}).
In the thermal-discrete approximation, we see from eq.\,(\ref{eq:T^cont.sum})
that
\beq
\tau_\rad\td (E_u) \approx \left[
\frac{1}{E_u}
\frac{8\pi}{h^3c^2} \int_0^{E_u} \frac{E^3 C_\abs(E)}{\exp(E/kT_u)-1}dE
\right]^{-1}
~~~
{\rm for}~u>1.
\eeq
Cooling times for PAH and silicate grains 
are plotted in Figs.\ \ref{fig:PAH_cooling_time}
and \ref{fig:sil_cooling_time}.
The interstellar starlight radiation field (ISRF) of
Mathis, Mezger \& Panagia (1983, hereafter MMP) has been assumed for $u_E$;
for this very dilute radiation field the correction for 
stimulated emission is
entirely negligible ($h^3c^3u_E/8\pi E^3 \ll 1$ for $E/hc \gtsim 10\cm^{-1}$).


We show $\tau_\rad$ evaluated using the ``exact-statistical'' method,
as well as using the ``thermal-discrete'' approximation
($\tau_\rad$ for the continuous cooling approximation is, by
construction, identical to that for the ``thermal-discrete'' approximation).

As illustrated in Figs.\,\ref{fig:PAH_cooling_time}-\ref{fig:sil_cooling_time},
$\tau_\rad(E)$ increases with grain size; 
for a given grain size, $\tau_\rad$ decreases rapidly with  
increasing vibrational energy $E$. 
The importance of single photon 
heating is apparent: very small grains emit most of their energy 
very rapidly (within seconds) following
absorption of an energetic photon and spend most of their time in or
close to the ground vibrational state.

The substantial difference between the results of the statistical
model and the thermal models at low $E$ is mainly due to the 
treatment of transitions to the ground state, which
are problematic in the thermal approximation
(see \S\ref{sec:thermal-discrete}). 
The jagged behavior of $\tau_\rad$ in the exact-statistical model 
result from 
the discreteness of the lowest energy states
(see eq.\,\ref{eq:model_mode_spec}) which affects 
the degeneracies (see eq.\,\ref{eq:T_lu}).
The discontinuity in $\tau_\rad$ computed in the thermal approximation
(Figures \ref{fig:PAH_cooling_time},\ref{fig:sil_cooling_time}) is due
to the discontinuity in the prescription (\ref{eq:T_u}) for the grain 
``vibrational temperature'' at $E=\hbar\omega_{20}$ 
(see Figs.\ \ref{fig:TvsE_PAH},\ref{fig:TvsE_sil}).
Note that with the present prescription for the temperature, our estimates
for $\tau_\rad$ in the thermal approximation are in reasonable agreement
with $\tau_\rad$ calculated in the ``exact-statistical'' method at
the lowest energies.
For C$_{4000}$H$_{1000}$ in Figure \ref{fig:PAH_cooling_time},
and $a=20\Angstrom$ in Figure \ref{fig:sil_cooling_time},
the discontinuity in $\tau_\rad$ at $\sim3\times10^4\cm^{-1}$ is due
to a transition from direct computation of the density of states to use
of a continuous Debye model.

\begin{figure*}[ht]
\centerline{
	\epsfig{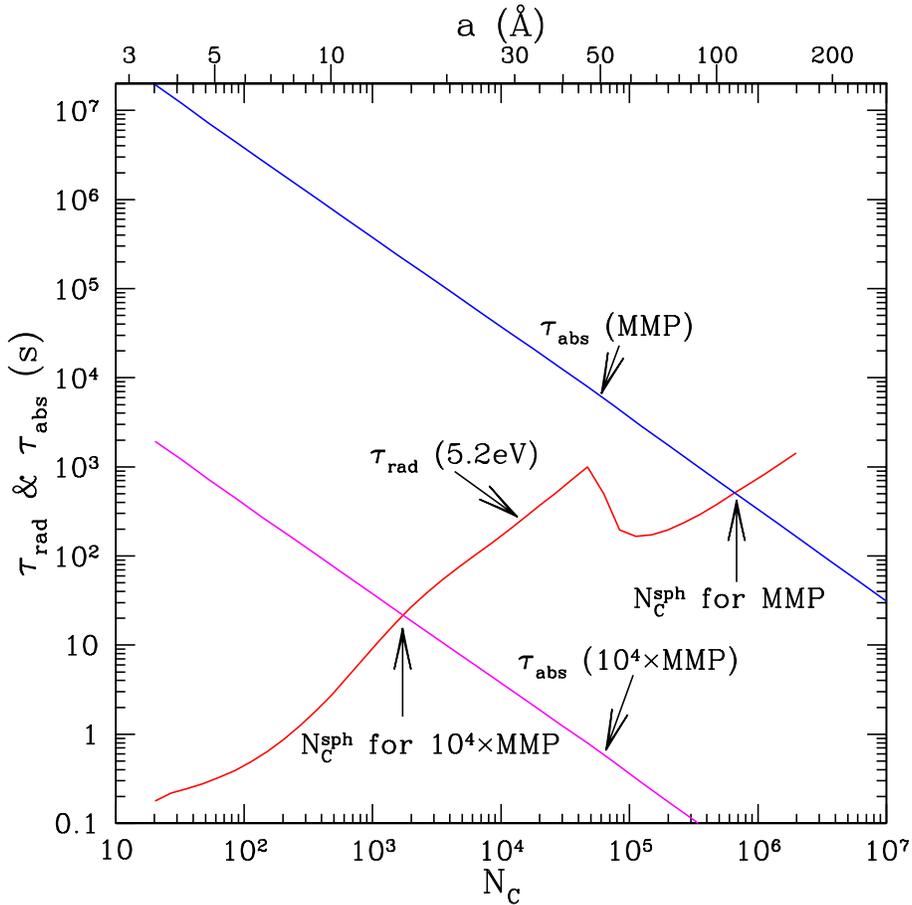}
	}
\figcaption{\footnotesize
	\label{fig:taurad_tauabs}
	Radiative cooling time $\tau_\rad$
	for a grain containing $E=5.2\eV$ of
	vibrational energy, and
	mean time $\tau_{\rm abs}$ between
	photon absorptions for the MMP ISRF,
	as a function of grain size, for carbonaceous and silicate
	grains.
	The decrease in $\tau_\rad$ as $N_\C$ increases above $\sim5\times10^4$
	is due to the assumed transition from PAH-like to graphitic
	optical properties at $a\approx 50\Angstrom$.
	}
\end{figure*}

The mean time $\tau_{\rm abs}$ between photon absorptions
for a grain is given by
\beq
\tau_{\rm abs}^{-1} \equiv 
\int_0^\infty C_{\rm abs}(h\nu)\frac{cu_\nu}{h\nu} d\nu ~~,
\eeq
In Figure \ref{fig:taurad_tauabs} we show $\tau_{\rm abs}$ for grains
in the MMP ISRF.
The mean
absorbed photon energy $\langle h\nu\rangle_{\rm abs}$ is
\beq
\langle h\nu \rangle_{\rm abs} \equiv
\tau_{\rm abs}\int_0^\infty C_{\rm abs}(h\nu) c u_\nu d\nu ~~.
\eeq
Ionized PAHs with $N_\C < 10^4$ in the MMP radiation field have
$\langle h\nu\rangle_{\rm abs}\approx5.2\eV$; neutral PAHs have a somewhat
higher value of $\langle h\nu\rangle_{\rm abs}$ due to their different
absorption cross section (see Figure \ref{fig:PAHcs}).

Also plotted in Figure \ref{fig:taurad_tauabs}
is the cooling time $\tau_\rad$ for grains containing
5.2~eV of thermal energy.
The critical size $N_{\rm sph}$ (and radius $a_{\rm sph}$) for
single-photon heating is determined by the condition
$\tau_\rad(N_{\rm sph})=\tau_{\rm abs}(N_{\rm sph})$.
For carbonaceous grains, we find 
\beq
N_{\C,\rm sph} \approx 
5\times10^5 \left(\frac{u}{u_{\rm MMP}}\right)^{-0.60}
~,~~ a_{\rm sph} \approx 
100\Angstrom \left(\frac{u}{u_{\rm MMP}}\right)^{-0.20}
~~.
\label{eq:N_sph}
\eeq
For grains with $N < N_{\rm sph}$ atoms, single-photon heating effects are
very important:
the radiative cooling time is
short compared to the interval between photon absorptions, and
the grains spend most of the time between photon absorptions
at energies $E\ll\langle h\nu\rangle_{\rm abs}$.
Grains with $N \gg N_{\rm sph}$
heat up until their energy content is large compared to
the energies of starlight photons, and individual photon absorptions
produce small fractional changes in the thermal energy content.
For such large grains, one may therefore estimate the average energy
content $\bar{E}$ 
by requiring a steady balance between cooling and heating:
\beq
\frac{\bar{E}}{\tau_\rad(\bar{ E})} =
\int d\nu ~c u_\nu C_{abs}(h\nu) ~~~.
\eeq

\section{Solution Method
	\label{sec:solution_method}}

\subsection{Discrete Cooling
	\label{sec:discrete-cooling-solution}
	}

If we define the diagonal elements of $\bT$ to be
\beq
\bT_{ii} = -\sum_{j\neq i}\bT_{ji}
\eeq
then equation (\ref{eq:Pdot}) becomes
\beq
\sum_{j=0}^{\Nb} \bT_{ij}P_j = 0 ~{\rm for}~ i=0,...,\Nb ~~.
\label{eq:Pdot2}
\eeq
Combining this with
the normalization condition $\sum_{j=0}^{\Nb} P_j=1$, we obtain
a set of $\Nb$ linear equations for the first $\Nb$ elements of 
$P_j$:
\beq
\sum_{j=0}^{\Nb-1}\left(\bT_{ij}-\bT_{i\Nb}\right)P_j = -\bT_{i\Nb}
~~~{\rm for}~i=0,...,\Nb-1 ~.
\label{eq:lineq}
\eeq
The remaining undetermined element $P_{\Nb}$ is obtained by
\beq
P_{\Nb} = -\left(\bT_{\Nb\Nb}\right)^{-1} \sum_{j=0}^{\Nb-1} \bT_{\Nb j}P_j
~~~.
\eeq
For small $\Nb$ we can directly solve 
eq. (\ref{eq:lineq}) for $P_j$ using
Gaussian elimination.
When $M\gtsim10^2$, however,
direct solution, requiring $O(\Nb^3)$ operations, is both slow
and numerically ill-behaved.
We therefore resort to iterative techniques,
and have tried both
the bi-conjugate gradient (BiCG) method 
(Press et al. 1992), 
and the stabilized bi-conjugate gradient (BiCGstab) method (see Sleijpen \&
Fokkema 1993 and references therein). We found the BiCG method to be quite
efficient even without preconditioning,
whereas the convergence rate of the BiCGstab method is dependent on the 
choice of preconditioner. 
   
Ideally, the adopted highest energy bin $E_{\Nb}$ for a given 
grain would be as large as possible, but large values of $M$ entail
heavy memory and cpu requirements.
In our calculations, we first
set $E_{\Nb}=13.6\eV$ and solve eq. (\ref{eq:lineq}) for $P_j$. 
If $P_{\Nb} < 10^{-14}$ then we use this $E_\Nb$ and solution $P_j$; otherwise
we increase $E_\Nb$ by a factor 1.5 and solve again, repeating until
we obtain a solution with $P_\Nb<10^{-14}$.

Small ($a\ltsim20$\AA) grains
exposed to the interstellar starlight background
are able to cool completely betweeen photon absorptions, 
and $E_\Nb=13.6\eV$ is sufficient. For larger grains, the interval between
photon absorptions is shorter and the cooling time at given $E$ is longer
(see Figs.\,\ref{fig:PAH_cooling_time} and \ref{fig:sil_cooling_time}),
so that the grain generally does not cool off to the ground state between
photon absorptions, and $E_{\Nb}$ must be
higher than 13.6eV. For a grain with $a\simgt 100$\AA\ only modest
cooling takes place between photon absorptions
and $P_j$ becomes strongly peaked around a
steady state ``equilibrium temperature''.

\subsection{Continuous Cooling
	\label{sec:continuous-cooling-solution}
	}

For the continous cooling approximation it is straightforward to
solve for the steady-state solution vector $P_j$.
As shown by
Guhathakurta \& Draine (1989),
if we define $X_j\equiv P_j/P_0$, then $X_0=1$ and the remaining $X_j$
may be obtained recursively:
\beq
X_j = \frac{1}{\bT_{j-1,j}\tc}
	\sum_{i=0}^{j-1} B_{j,i} X_i 
~~~,~~~
B_{j,i}\equiv\sum_{u=j}^{M}\bT_{u,i}\tc
\eeq
The recursion relation $B_{j,i}=\bT_{j,i}+B_{j+1,i}$ can be used
to generate the $B_{j,i}$ in $O(M^2)$ operations, and the
$X_j$ can then be computed in $O(M^2)$ operations.
Once the $X_j$ are determined for $j=0,...,M$, one need only renormalize:
\beq
P_j = \frac{X_j}{\sum_{j=0}^M X_j}
~~~.
\eeq
Since $M\gtsim100$ is required to have suitably narrow energy bins
(we often use $M\approx 500$),
this $O(M)$ procedure is {\it much} faster than the BiCG iterative method
required if
discrete cooling is included.

\section{Infrared Emission Spectrum
	\label{sec:IR_emission}}

Having obtained the steady-state energy distribution $P_j$,
we wish to calculate the resulting infrared emission spectrum.
Let $F_\nu d\nu$ be the time-averaged power per steradian 
per grain radiated into
frequency interval $[\nu,\nu+d\nu]$.
Then the ``exact statistical'' treatment gives
\beq
\nu F_\nu = \frac{2 h\nu^4}{c^2} C_{\abs}(\nu)
\left[
{\sum_{u}}^\prime P_u 
\sum_{l=0}^{u-1}
\frac{g_l}{g_u}\Delta E_u
	G_{ul}(h\nu)
+
{\sum_{u}}^{\prime\prime} P_u \left(1 - \frac{h\nu}{\Delta E_u}\right)
\right]
\left[1+\frac{\lambda^3}{8\pi}u_E\right]
\label{eq:nuFnu}
\eeq
The single-primed summation -- the contribution of ``interbin'' transitions
to the emission -- is limited to levels
$u$ with $E_{u,\max}> h\nu$.\footnote{%
	Note that only a few terms from $\sum_l^\prime$ have
	$G_{ul}\neq0$ -- energy levels $k\leq l\leq m$ where
	$E_{u,\min}-h\nu \in [E_{k,\min},E_{k,\max}]$ and
	$E_{u,\max}-h\nu \in [E_{m,\min},E_{m,\max}]$.
	Therefore eq.\ (\ref{eq:nuFnu}) requires only $O(M)$ 
	[rather than $O(M^2)$] operations per frequency $\nu$.
	}
The double-primed summation -- the contribution of ``intrabin'' transitions --
is limited to levels with $\Delta E_u > h\nu$.

In the thermal approximation, equation (\ref{eq:nuFnu}) becomes
(see Appendix \ref{app:spectrum})
\beq
\nu F_\nu \approx \frac{2 h\nu^4}{c^2} C_{\abs}(\nu)
\left[
{\sum_{u}}^\prime \frac{P_u}{\exp(h\nu/kT_u)-1}
\right]
\left[1+\frac{\lambda^3}{8\pi}u_E\right] ~~~,
\label{eq:nuFnu_thermal}
\eeq
where the single-primed summation is now limited to levels $u$ with
$E_u>h\nu$.

\section{Results
	\label{sec:results}}

We have calculated the (vibrational) energy probability distribution $P_j$
for PAHs excited by the MMP ISRF using
the above-described methods: 
the {\it exact-statistical} model (\S\ref{sec:statistical_treatment}); 
the {\it thermal-discrete} model (\S\ref{sec:thermal-discrete}); 
and the {\it thermal-continuous} model (\S\ref{sec:thermal-continuous}).

\begin{figure*}[ht]
\centerline{\epsfig{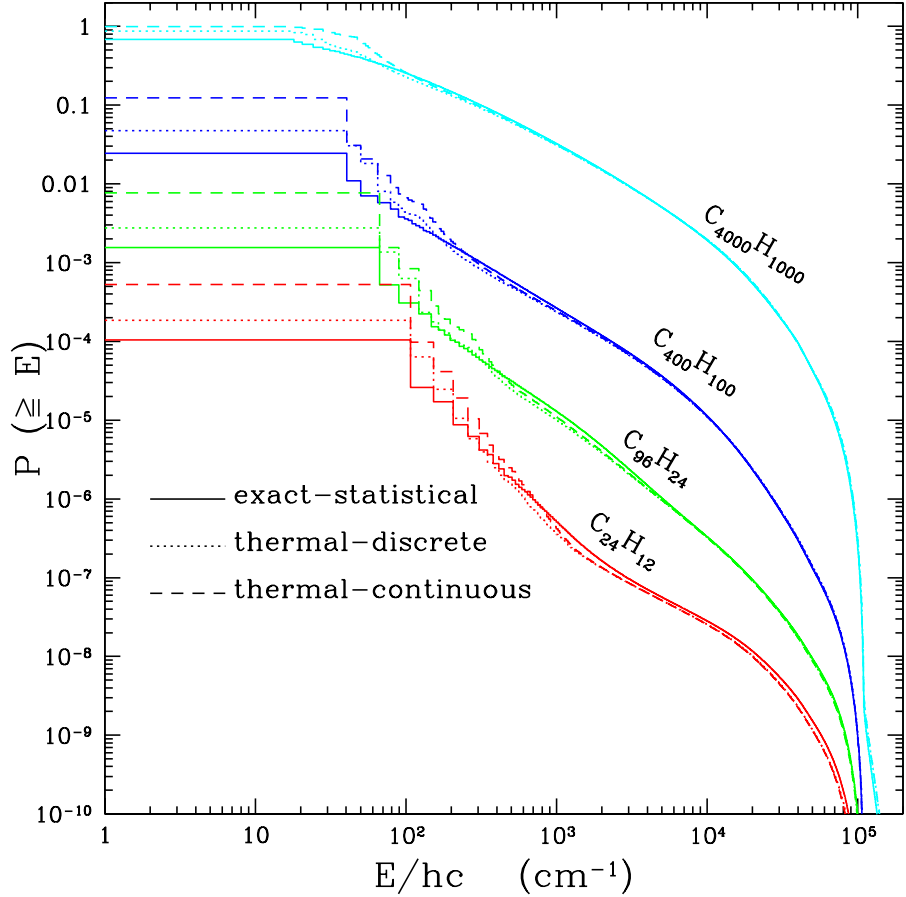}}
\figcaption{\footnotesize
	\label{fig:PE_distrib}
	The cumulative energy probability distributions for PAHs
	computed using the 
	{\it exact-statistical} model,
	the {\it thermal-discrete} model, 
	and the {\it thermal-continuous} model.
        Note that the lowest energy state ($E=0$), not shown here,
        has $P(E\geq 0)=1$.
        }
\end{figure*}
\begin{figure*}[ht]
\centerline{\epsfig{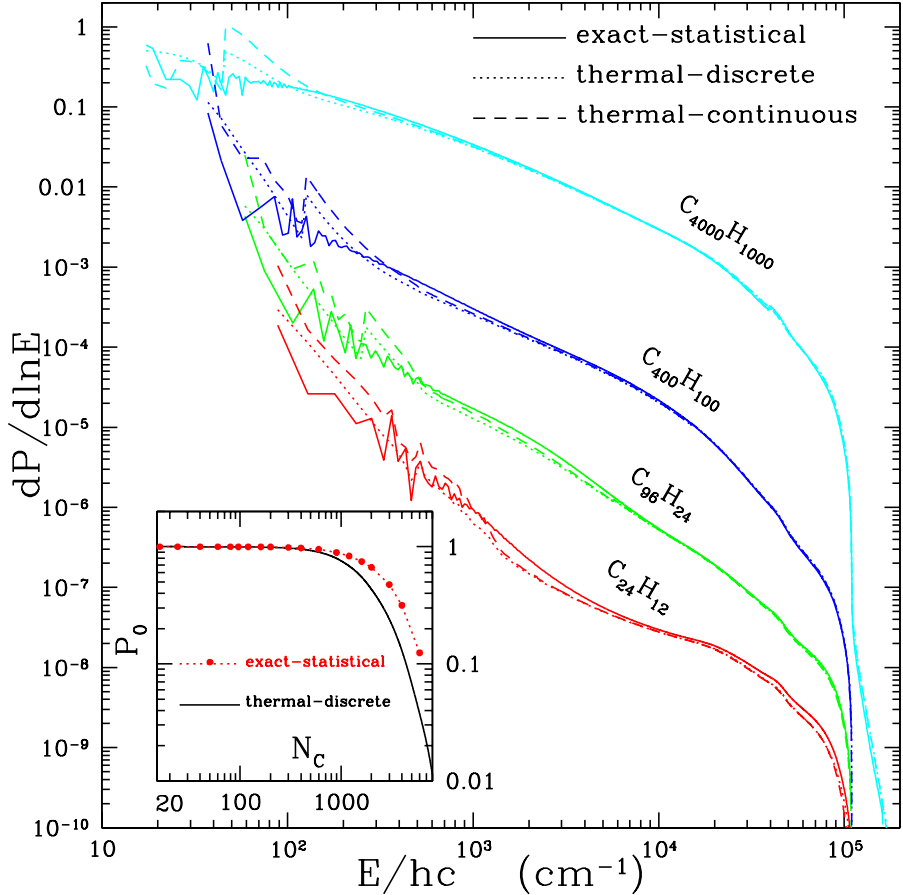}}
\figcaption{\footnotesize
        \label{fig:PE_diff_distrib}
        The vibrational energy distribution 
        for excited vibrational states of ionized PAHs in the MMP ISRF, 
	computed using the
	{\it exact-statistical} model,
        the {\it thermal-discrete} model, 
	and the {\it thermal-continuous} model.
        The inset shows the probability $P_0$ of being in the
	ground state.
         }
\end{figure*}

In Figure \ref{fig:PE_distrib} 
we present the {\it cumulative} energy probability distributions
for selected PAHs obtained from the exact-statistical model,
the thermal-discrete model, and the 
thermal-continuous model. 
Figure \ref{fig:PE_diff_distrib} shows the differential probability 
distributions as a function of grain energy $E$ for 
these same grain sizes.

The inset in Figure \ref{fig:PE_diff_distrib} shows the probability of being
in the vibrational ground state as a function of grain size, calculated for the
exact-statistical model and the
thermal-discrete model.
As already illustrated in Figure \ref{fig:PE_distrib} 
and Figure \ref{fig:PE_diff_distrib}, 
the probability of being
in the ground state is very large for small grains:
for example, for the MMP radiation field, 
grains with $N\ltsim4000$ spend most of their time at $E=0$
[the exact-statistical model gives $P_0= 0.975$ for $N=400$,
and $P_0=0.316$ for $N=4000$].
The sharp drop at 13.6\,eV ($E/hc=1.1\times 10^5$\,cm$^{-1}$)
is due to the 
radiation field cutoff at 912\,\AA\ and to the fact that multiphoton events
are rare. 
The jagged structure at low energies in Figure \ref{fig:PE_diff_distrib} 
is
due to the quantization of the lowest energy states. These structures are
less prominent in the thermal models because the statistical model involves
the degeneracies of each ``bin'' (see eq.\,\ref{eq:g_j}) 
which depend on the detailed distribution of energy states 
(see, e.g., Figure \ref{fig:dens_states}).

The energy distributions $P(E)$ found using the thermal-discrete model 
and the thermal-continuous model are both
in good overall agreement with the results of the exact-statistical
calculation.

It is not surprising that all three models are in good 
agreement at high energies 
since the transition rate from a very high energy state to the
ground state is negligibly small,
and the number of excited degrees of freedom is large enough that a thermal
treatment is appropriate. The agreement becomes better with 
increasing grain size because the lowest vibrational states (i.e., the first 
few excited states) become closer and approach a ``continuum'', and 
the gap between the ground state and the first excited state becomes
smaller. 

\begin{figure*}[ht]
\centerline{\epsfig{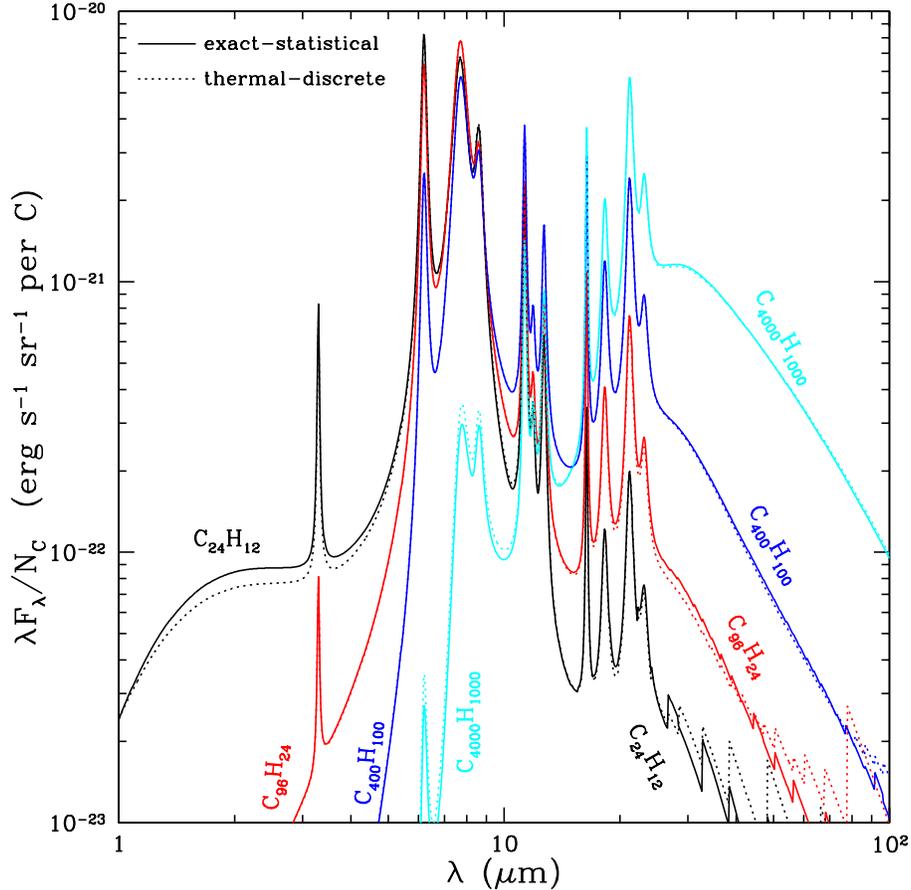}}
\figcaption{\footnotesize
        \label{fig:IR_stat_exact}
        IR emissivities (per C atom) for selected ionized PAHs
	in the MMP ISRF
        calculated using the {\it exact-statistical} 
        and {\it thermal-discrete} models.
	}
\end{figure*}
\begin{figure*}[ht]
\centerline{\epsfig{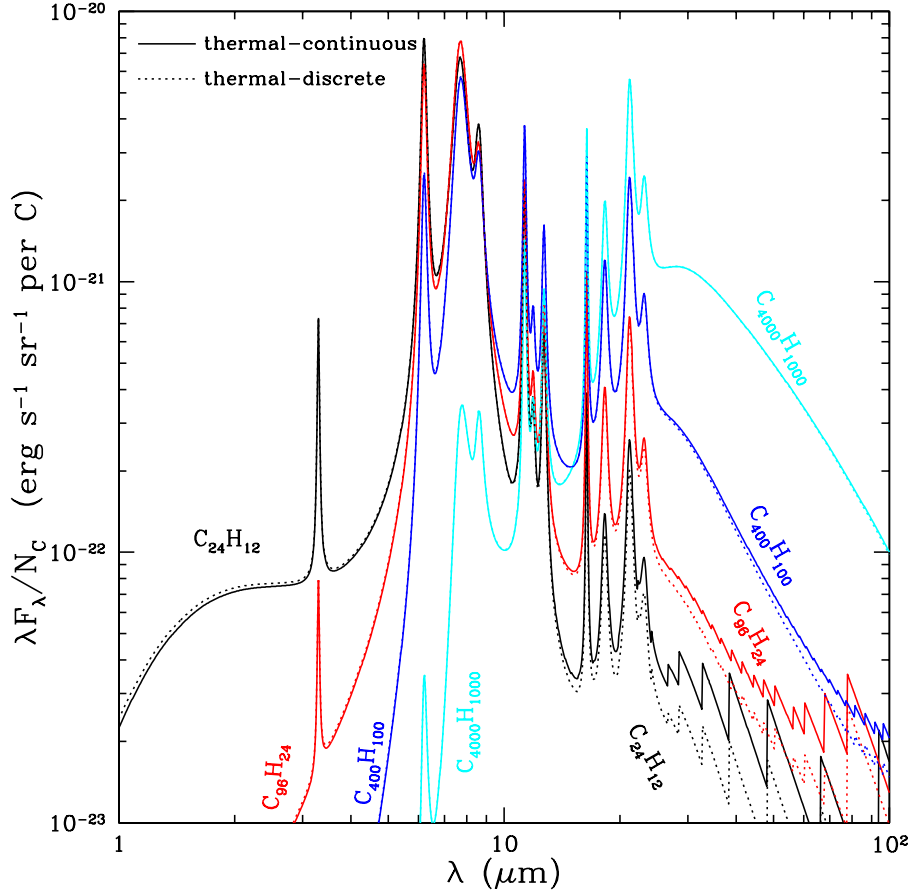}}
\figcaption{\footnotesize
        \label{fig:IR_discrete_continuous}
        IR emissivities (per C atom) for ionized PAHs of
        calculated using the {\it thermal-discrete}
        and {\it thermal-continuous} cooling models.
	}
\end{figure*}

The resulting IR emission spectra are displayed in 
Figs.\,\ref{fig:IR_stat_exact} and \ref{fig:IR_discrete_continuous}.
As expected (from the energy distribution functions plotted in 
Figs.\,\ref{fig:PE_distrib} and 
\ref{fig:PE_diff_distrib}), 
the IR spectra of the thermal-discrete model are almost identical to
those of the exact-statistical model, even for grains as small as
$N_\C=24$.
The thermal-continuous cooling model also results in spectra which are
very close to those computed using the exact-statistical model.
The discrepancies are mainly at long 
wavelengths; this can be understood from the energy probability functions
and would not affect the interstellar IR emission spectrum modeling.
 
The statistical model spectra are generally very close to those of the 
thermal models (Figs.\,\ref{fig:IR_stat_exact} and 
\ref{fig:IR_discrete_continuous}), 
except for the sawtooth features 
at long wavelengths. These sawtooth features are due to our treatment of 
transitions from the lower excited energy bins to the ground state and
first few excited states.
However, we stress
that the overall spectra are quite similar for all three models --
the differences involve only a negligible fraction
of the total emission for any given grain size, 
and thus would be unimportant when modeling the
overall interstellar IR emission spectrum.

\section{Discussion
	\label{sec:discussion}}

\subsection{Band Ratios}

From
Figs.\ \ref{fig:IR_stat_exact} and \ref{fig:IR_discrete_continuous}
it is evident
that the relative strength of the different PAH emission bands
(3.3, 6.2, 7.7, 8.6, 11.3$\micron$) is a strong function of the PAH size:
small PAHs radiate strongly at 6.2 and 7.7$\micron$, while larger PAHs
emit most of their power at increasingly long wavelengths.

\begin{figure*}[ht]
\begin{center}
	\hspace*{-15cm}
	\begin{rotate}{-90}
	\epsfig{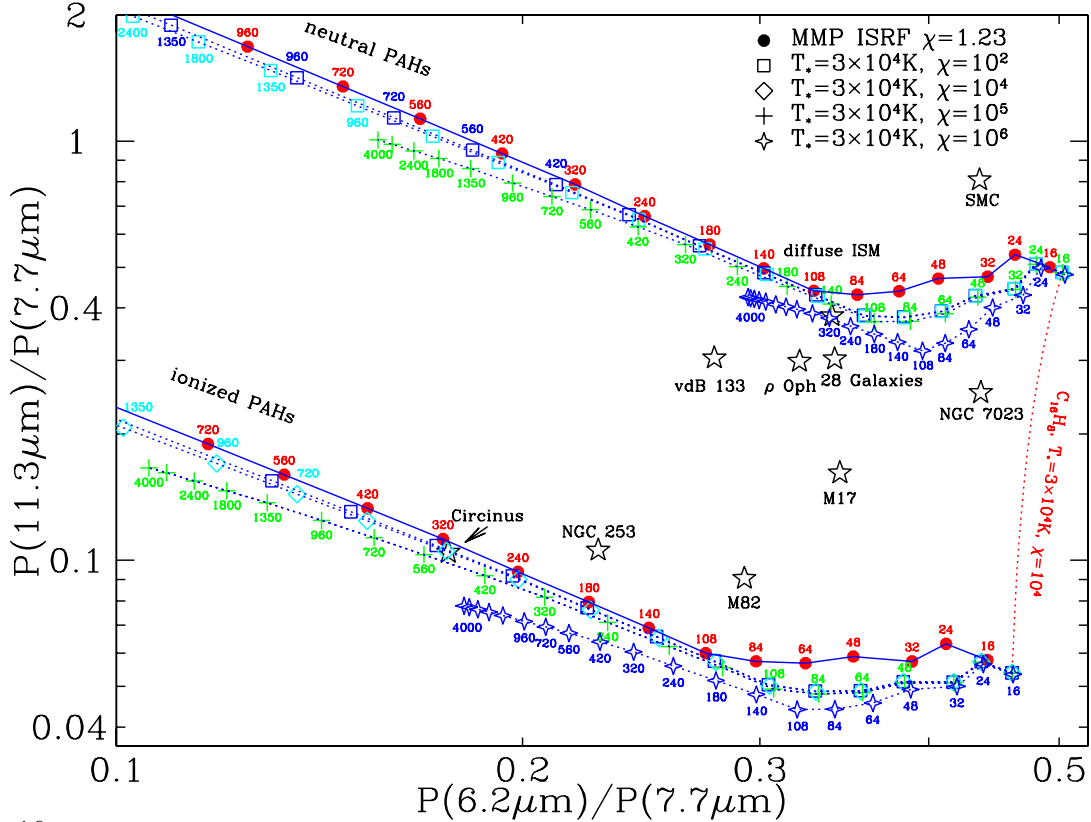}
	\end{rotate}\vspace*{10cm}
\end{center}
\figcaption{\footnotesize
	\label{fig:band_ratios_0}
        Relative strengths of three PAH emission features  
        (6.2 and 7.7$\mu$m C-C stretching modes, and 11.3$\mu$m 
        C-H out-of-plane bending mode) for neutral and ionized PAHs,
	labelled by the number of carbon atoms. 
	Results are shown for grains illuminated by the MMP spectrum
	with $\chi=1.23$ and $123$, and for a $3\times10^4\K$
	blackbody spectrum (cutoff at 912$\Angstrom$) with
	$\chi=10^2$, $10^4$, $10^5$, and $10^6$.
        Also shown (stars) are observed band ratios
        for the diffuse ISM,
        the reflection nebulae NGC 7023 and vdB 133,
        the M17 PDR,
	the $\rho$ Oph molecular cloud;
	the starburst galaxies M82 and NGC 253 (St\"urm et al.\ 2000);
	the Seyfert 2 galaxy in Circinus;
        the average of 28 normal galaxies;
	and a quiescent molecular cloud in the SMC.
	(See text for citations.)
	The dotted line connecting the neutral and ionized points
	for $N_\C=16$ (in a PDR) illustrates the variation in the
	relative strengths of the 6.2, 7.7, and 11.3$\micron$ features
	in a PDR
	as the PAH ion fraction varies from 0 to 1.
	}
\end{figure*}

The 6.2, 7.7, and 11.3$\micron$ features are prominent in many
astronomical spectra.
Let $P(\lambda_0)$ be the power radiated in the feature with central
wavelength $\lambda_0$: for a Drude profile,
$P(\lambda_0)=(\pi/2)(\Delta P_\lambda)_0 ({\rm FWHM})_{\lambda_0}$,
where $(\Delta P_\lambda)_0$ is the peak contribution of the feature to the
emission, and $({\rm FWHM})_{\lambda_0}$ is the full-width at half-maximum.
In
Figure \ref{fig:band_ratios_0} we show how the relative strengths of these
emission bands vary depending upon the size and charge state of the PAHs, 
and upon the starlight intensity $u_E$,
characterized by $\chi$ or $G_0$.\footnote{
	$\chi$ is the intensity at 1000$\Angstrom$ relative to
	the estimate of Habing (1968);
	$G_0$ is the ratio of the 6--13.6~eV energy density
	to the estimate of Habing (1968).
	The MMP spectrum has $G_0=0.923\chi$; a
	$3\times10^4\K$ blackbody has $G_0=0.602\chi$.
	} 
Neutral PAHs have large ratios of 
$P(11.3\micron)/P(7.7\micron)$,
while PAH ions have much smaller values.
For both neutrals and ions, there is a regular progression of
$P(6.2\micron)/P(7.7\micron)$
to smaller values with increasing $N_\C$.
For a given PAH neutral or ion,
the emission band ratios are essentially 
independent of $\chi$ for $\chi < 3\times10^4(10^3/N_\C)^{5/3}$
[see eq.\ (\ref{eq:N_sph})]; for larger values of $\chi$,
multiphoton heating effects begin to be evident.
For $\chi=10^6$, multiphoton effects are evident even for
$N_\C=10^2$ [consistent with eq.\ (\ref{eq:N_sph})].

\begin{figure*}[ht]
\begin{center}
	\hspace*{-15cm}
	\begin{rotate}{-90}\epsfig{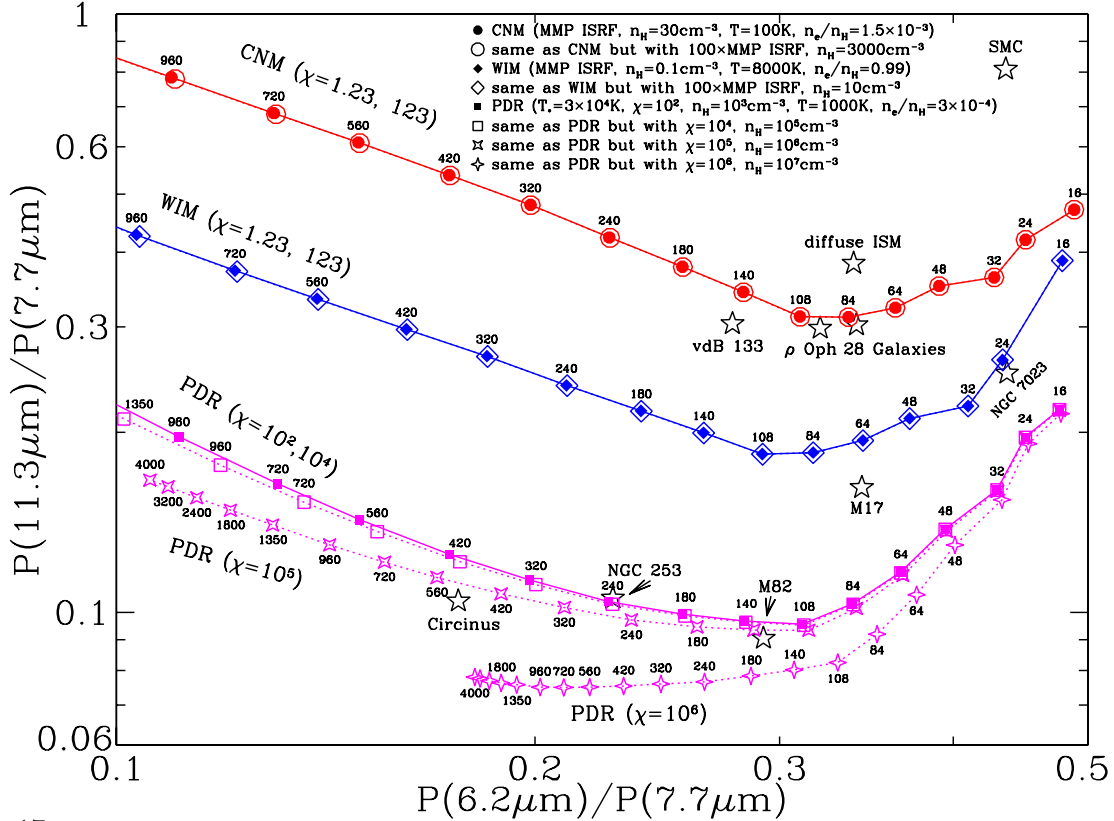}
	\end{rotate}\vspace*{10cm}
\end{center}
\figcaption{\footnotesize
	\label{fig:band_ratios_1}
        Relative strengths of three PAH emission features  
        (6.2 and 7.7$\mu$m C-C stretching modes, and 11.3$\mu$m 
        C-H out-of-plane bending mode); PAHs are labelled by the
        number of carbon atoms. 
	Results are shown for 3 charging conditions: CNM, WIM, and PDR
	(see text).
	Circles:
        grains in CNM charging conditions,
	for $G_0=1.14$ and 114;
	diamonds: grains in WIM charging conditions,
	for $G_0=1.14$ and 114;
	squares: grains in
	PDR conditions,
	for $\chi=10^2$, $10^4$, $10^5$, and $10^6$.
	Stars represent observations as in Fig.\ \ref{fig:band_ratios_0}.
	}
\end{figure*}
\begin{figure*}[ht]
\begin{center}
	\epsfig{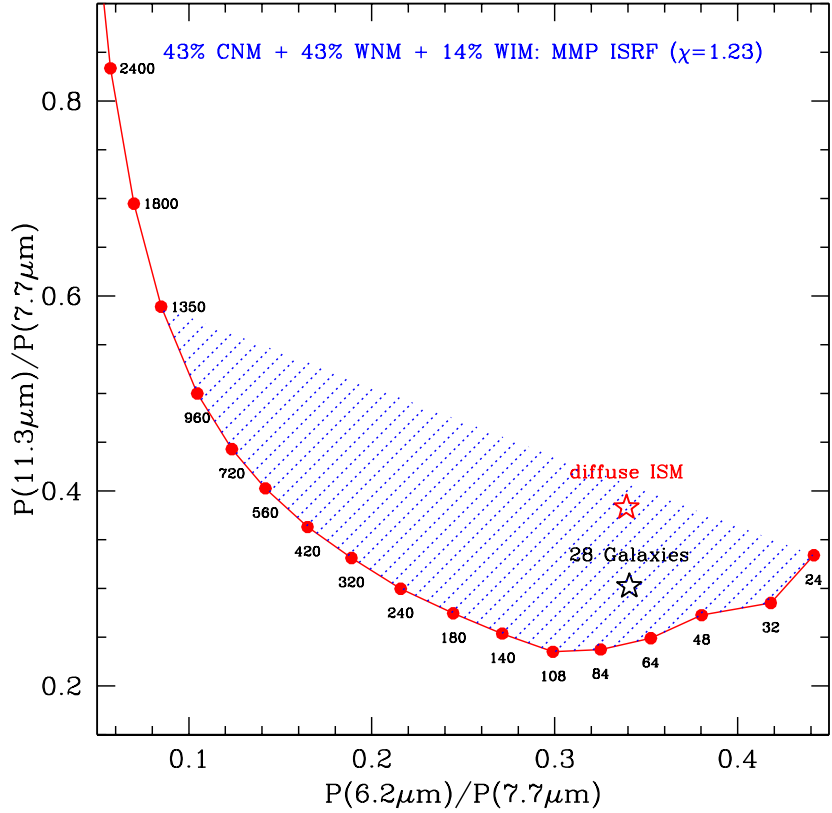}
\end{center}
\figcaption{\footnotesize
	\label{fig:band_ratios_2}
	Same as Fig.\ \ref{fig:band_ratios_0} but for a weighted sum of
	emission from grains in the cold neutral medium (CNM),
	grains in the warm neutral medium (WNM), and grains in
	the warm ionized medium (WIM), heated by the MMP estimate of
	the ISRF.
	Band ratios within the shaded region can be obtained by
	appropriate mixtures of PAHs with $20\leq N_\C\leq 1000$.
	Observed band ratios for the diffuse ISM (Onaka et al. 1996;
	Mattila et al. 1996) and the average spectrum of 28 normal
	spiral galaxies (Helou et al. 2000) fall within this allowed
	region.
	}
\end{figure*}

In Figure \ref{fig:band_ratios_0} we plot the observed ratios
of 
$P(11.3\micron)/P(7.7\micron)$
and 
$P(6.2\micron)/P(7.7\micron)$
for
(1) the diffuse ISM (Onaka et al.\ 1996);
(2) the average spectrum for a sample of 28 spiral galaxies
(Helou et al.\ 2000);
(3) the reflection nebula at the surface of the 
$\rho$ Oph molecular cloud (Boulanger et al.\ 1996);
(4) the reflection nebula vdB 133,
illuminated by F5Iab and B7II stars (Uchida et al.\ 1998);
(5) the reflection nebula/photodissociation region (PDR) 
NGC 7023, illuminated by a B2.5V star 
(Cesarsky et al. 1996a);
(6) the M17 PDR (Cesarsky et al.\ 1996b);
(7) the starburst galaxies M82 and NGC 253 (St\"urm et al.\ 2000);
(8) the Seyfert 2 galaxy in Circinus (St\"urm et al.\ 2000);
and
(9) the quiescent molecular cloud SMC B1\#1 in the Small Magellanic Cloud 
(Reach et al.\ 2000).
With the exception of SMC B1\#1 (which we discuss below),
the observed band ratios fall 
between the neutral and ionized PAH tracks in 
Figure \ref{fig:band_ratios_0},
suggesting that the observed spectra can be reproduced
by an appropriate mixture of PAH sizes, with an appropriate
ionized fraction.
The observed band ratios for the
Milky Way ISM, $\rho$ Oph, 28 normal galaxies,
and the reflection nebulae vdB 133 and NGC 7023
could be reproduced
with a relatively low ionized fraction, but larger ionized fractions are
required for the M17 PDR, M82, NGC 253, and the Circinus galaxy.

\subsection{SMC B1\#1}

The quiescent molecular cloud SMC B1\#1 observed by Reach et al.\ (2000)
falls outside the region bounded by the neutral and ionized
tracks in Figure \ref{fig:band_ratios_0}.
Reach et al.\ point out that SMC B1\#1 has a 11.3/7.7 band ratio which
differs
significantly from that for Milky Way objects, including the $\rho$ Oph
reflection nebula (which one might expect SMC B1\#1 to resemble).
Reach et al.\ suggest that the enhanced emission in the
11.3$\micron$ C-H bending mode 
might be
due to increased hydrogenation of the PAHs in the SMC, where the gas phase
H/C ratio is a factor of 10 higher than in the Milky Way.
Given that Milky Way PAHs with $\gtsim$25 C atoms
are expected to be essentially fully hydrogenated
(Tielens et al.\ 1987; Allamandola, Tielens, \& Barker 1989; 
Allain, Leach, \& Sedlmayr 1996),
it is not clear that greater hydrogenation would be expected in the SMC.
As discussed in Paper II, our model calculations assume that
interstellar neutral and ionized PAHs have the 
11.3/7.7$\micron$ band strength ratio 
{\it reduced}
by a factor of 2 relative to PAHs which have been studied in the lab --
enhancement of the 7.7$\micron$ feature appears to be necessary to
reproduce the low 11.3/7.7$\micron$ band ratios observed in objects such
as NGC 253 or M82, and (with the possible 
exception of SMC B1\#1) the reduced 11.3/7.7$\micron$
band strength ratio appears to be consistent with astronomical observations.

While SMC B1\#1 appears to fall ``outside'' the ``allowed'' region in
Figures \ref{fig:band_ratios_0}--\ref{fig:band_ratios_2}, 
we show elsewhere (Li \& Draine 2001b) that the present model can
reproduce the observed line ratios by adding emission from
a population of larger PAHs -- with $N_\C\approx 3000$ -- 
which radiate strongly in
the 11--13$\micron$ region, but produce little emission in the
6--8$\micron$ region.
These larger PAHs would radiate strongly at 15--50$\micron$;
SIRTF will be able to test this hypothesis.

\subsection{PAH Ionization}

The fraction of PAHs which are charged will depend upon the electron
density $n_e$, temperature $T$, and the starlight intensity.
We consider three different charging environments:
\begin{itemize}
\item ``Cold neutral medium'' (CNM;
$n_e=0.045\cm^{-3}$, $T=100\K$, and the MMP ISRF spectrum
with $\chi=1.23$, $G_0=1.14$)
and with both radiation field and
gas density increased by $10^2$;
\item ``Warm ionized medium'' (WIM;
$n_e=0.1\cm^{-3}$, $T=8000\K$, and the MMP ISRF with $\chi=1.23$),
and with both radiation field and gas density
increased by $10^2$;
\item Photodissociation region conditions 
(PDR;
$n_e/n_\rmH=3\times10^{-4}$, $T=10^3\K$, and
a $3\times10^4\K$ blackbody cut off at 912$\Angstrom$
with $\chi/n_\rmH=0.1\cm^3$) for $\chi=10^2$, $10^4$, $10^5$, and $10^6$.
While PDRs come in a range of conditions, recent observations and
modeling suggest that at least some PDRs include 
an extended region with $T\approx10^3\K$, and with 
$\chi/n_\rmH\approx0.1\cm^3$, as in a recent model for the 
NGC 2023 PDR (Draine \& Bertoldi 2000).
\end{itemize}
For each set of conditions, we consider PAHs with $N_\C\geq16$ C atoms,
and include emission from both neutral and ionized forms, with the
neutral fraction calculated assuming a balance between collisional
charging by electrons and photodetachment and photoionization
(Weingartner \& Draine 2001b).
The charging of small grains is largely a function of the quantity
$\chi\sqrt{T}/n_e$ (Bakes \& Tielens 1994; 
Weingartner \& Draine 2001b).
For the CNM, WIM, and PDR conditions considered here,
$\chi\sqrt{T}/n_e$ = 270, 1100, and $10^4$ $\cm^3\K^{1/2}$, so we expect
grains of a given size to be increasingly positively charged as
we move from CNM to WIM to PDR.

The ``diffuse ISM'' and ``28 Galaxies''
points fall between the CNM and WIM tracks
in Figure \ref{fig:band_ratios_0}, implying that one could
reproduce the overall emission from normal spiral galaxies by
a weighted sum of interstellar conditions ranging from CNM to WIM,
with a mixture of grain sizes such that the 6.2 -- 11.3$\micron$
emission is dominated by grains with $20\ltsim N_\C \ltsim 500$.

The observed flux ratios for NGC 7023 and M17 fall between
the ``WIM'' and ``PDR'' tracks in Figure \ref{fig:band_ratios_1}.
The observed flux ratios for the starburst galaxies
M82 and NGC 253, as well as the Circinus Seyfert 2 galaxy, 
fall close to the ``PDR'' tracks,
consistent with the expectation that the spectra of these
galaxies includes a substantial contribution from PDRs.

It is important to note that the ``PDR'' models in 
Figure \ref{fig:band_ratios_1}
have charging conditions such that a substantial fraction of the
small grains are still neutral: 22\% of the $N_\C=100$ grains are
neutral.
Larger charged fractions -- and smaller values of 
$P(11.3\micron)/P(7.7\micron)$
-- can be achieved by increasing the value of $\chi/n_\rmH$.
For example, if $\chi/n_\rmH$ is increased by a factor 2, only
10\% of the $N_\C=100$ grains are neutral.

In Figure \ref{fig:band_ratios_2} we show the emission expected from
grains of various sizes, with 43\% in the CNM, 14\% in the WIM, and
43\% in the WNM ($n_e=0.03\cm^{-3}$, $T=6000\K$, MMP ISRF,
$\chi\sqrt{T}/n_e=3200$).
The shaded region shows the flux ratios which could be obtained by
grain mixtures containing only grains $20<N_\C<1350$
(from Figure \ref{fig:IR_discrete_continuous} it is apparent that grains
with $N_\C>10^3$ are not likely to contribute significant emission
at $\lambda \leq 11.3\micron$).
The observed flux ratios for the diffuse ISM in the Milky Way, and
for the average of 28 spiral galaxies, both fall in the shaded region,
suggesting that with an appropriate size distribution, one might
be able to reproduce the observed emission.
The question of size distributions will be addressed in a
future paper (Li \& Draine 2001a), where it will be shown that 
the observed infrared emission can be approximately reproduced by
grain
size distributions similar to those invoked to explain the
observed microwave emission from interstellar dust 
(Draine \& Lazarian 1998a,b).

\section{Summary
	\label{sec:summary}}
We have presented a method for calculating the infrared
emission from dust grains, including very small PAH molecules or
silicate clusters.
The principal results of this paper are the following:
\begin{enumerate}
\item The vibrational mode spectrum and vibrational density of states
	of very small grains are discussed in 
	\S\ref{sec:vibrational_mode_spectrum}, where we
	explicitly consider both polycyclic aromatic hydrocarbons and
	silicate grains.
\item We use the adopted vibrational mode spectrum to calculate 
	the specific heat of carbonaceous and
	silicate grains; the results
	are in agreement with experimental
	specific heats for graphite and amorphous silicates
	(Figure \ref{fig:spec_heats}).
\item Using estimates for the absorption cross sections of carbonaceous
	and silicate grains, we obtain (in \S\ref{sec:statistical_treatment})
	radiative transition rates for emission from
	vibrationally-excited states of carbonaceous and silicate grains.
	We refer to this as the ``exact-statistical'' treatment.
\item We discuss (in \S\ref{sec:thermal_treatment})
	how the ``thermal approximation'' can be used to
	estimate rates for spontaneous emission.
	With a judicious estimate for the ``temperature'', the
	thermal approximation provides transition rates which are
	close to those obtained from the ``exact-statistical'' treatment.
	Therefore level populations estimated using the thermal approximation
	are close to the actual level populations given by the
	``exact-statistical'' treatment.
\item We discuss the ``continuous cooling'' approximation for modeling
	the deexcitation of the grain.
	We show that level populations
	and emission spectra computed using this approximation are 
	sufficiently accurate for most astrophysical applications,
	even for grains containing as few as $\sim30$ atoms.
\item Methods for numerical solution for the vibrational energy distribution
	are presented in \S\ref{sec:solution_method}, and the numerical
	advantages of the continuous cooling approximation are stressed.
\item Relative strengths of the 6.2, 7.7, and 11.3$\micron$ features
	depend on the grain size, on the charging conditions, and on the
	starlight intensity.  Observed 6.2/7.7 and 11.3/7.7 band ratios
	for the diffuse ISM in the Milky Way, for the $\rho$ Oph and
	vdB133 reflection 
	nebulae, and for the
	28 ``normal'' galaxy sample of Helou et al. (2000),
	appear consistent with
	grains in the diffuse ISM heated by the average starlight background.
	The band ratios observed for the starburst galaxies
	M82 and NGC 253, for the Seyfert 2 ``Circinus galaxy'',
	for the M17 PDR, and for the NGC 7023 PDR
	appear to require conditions close to our ``PDR'' conditions,
	with $\chi/n_\rmH\approx 0.05 - 0.1\cm^3$.
\item The unusual band strengths observed for the quiescent molecular cloud
	SMC B1\#1 in the SMC (Reach et al.\ 2000) cannot be reproduced by
	a mixture of small $N_\C\ltsim 10^3$ PAHs.
	Either the PAHs in this cloud differ from Milky Way PAHs in their
	properties, or have a different size distribution.
\end{enumerate}
The present paper has concentrated on the statistical mechanics of the
stochastic heating process, and on the resulting emission spectra as a
function of grain size.
In a separate paper (Li \& Draine 2001a) we calculate the emission from
grain models with realistic size distributions, and show that the observed
infrared emission from interstellar dust can be reproduced by a grain model
with $\sim$15\% of the interstellar carbon in PAH particles containing
less than $10^3$ C atoms.

\acknowledgements
We thank L.J. Allamandola, J.M. Greenberg, and J.C. Weingartner for 
helpful discussions.
We thank D.M. Hudgins and L.J. Allamandola for providing us with their
PAH database.
We thank R.H. Lupton for availability of the SM plotting package.
This research was supported in part by NSF grant AST-9619429,
and NASA grant NAG5-7030.

\appendix
\section{PAH Geometry\label{app:geom}}

Small PAHs are expected to be planar, with $D\propto N_\C^{1/2}$, but
large PAHs are expected to be approximately spherical, with
$D\propto N_\C^{1/3}$.
The size at which the transition from planar to spherical geometry occurs
is uncertain.
We will assume that PAHs have a planar geometry up to
$N_\C=54$
(e.g., the pericondensed species circumcoronene C$_{54}$H$_{18}$).
We assume that beyond $N_\C=54$ PAHs consist of multiple sheets.
Beyond $N_\C=102$ 
(e.g., C$_{54}$H$_{18}$ plus one coronene C$_{24}$H$_{12}$ on either side)
we assume spherical geometry, with 
$D\propto N_\C^{1/3}$.

\section{\label{app:energy_bins}Energy Bins}

We specify the energy bins so that the total number $\Nb$ 
of bins is manageable ($300\ltsim \Nb \ltsim 1500$),
the number of vibrational states per bin is at least 2 for each of
bins 3--10, the bin widths vary smoothly for $j>10$, and the last
bin is at a high enough energy that its population $P_\Nb < 10^{-14}$
(see \S\ref{sec:discrete-cooling-solution}
below).

Let $\omega_j, j=1,...,3(N_a-2)$ be the fundamental vibrational
modes (see \S\ref{sec:vibrational_mode_spectrum}), 
in order of increasing frequency.
Bin 0 is the ground state: $E_{0,\min}=E_{0,\max}=E_0=0$, to
which we assign a ``width'' $\Delta E_0=0$.
Bins 1 and 2 contain the first two vibrational modes, and
bins 3 through $10$ each contain two normal modes:
\begin{eqnarray}
E_{1,\min}&=&\frac{3}{2}\hbar\omega_1-\frac{1}{2}\hbar\omega_2\\
E_{j,\max}=E_{j+1,\min}&=&
\frac{1}{2}\left(\hbar\omega_j+\hbar\omega_{j+1}\right)
~~~{\rm for~}j=1,2\\
E_{j,\max}=E_{j+1,\min}&=&
\frac{1}{2}\left(\hbar\omega_{2j-2}+\hbar\omega_{2j-1}\right)
~~~{\rm for~}3\leq j\leq 10
\end{eqnarray}
For $j=11,...,K$ we use uniform intervals $\Delta E$, and for
$j\geq K$ uniform intervals in $\ln E$:
\beq
E_{j,\max}=E_{10,\max}+(j-10)(E_{10,\max}-E_{9,\max}) 
~{\rm for}~ 10\leq j \leq K
\eeq
\beq
E_{j,\max}=E_{K,\max}\left(\frac{E_{K,\max}}{E_{K-1,\max}}\right)^{j-K}
{\rm for~}K\leq j\leq\Nb ~~~,
\eeq
$K$ is chosen so that the total number of bins $\sim500$.

\section{\label{app:T_fi}Transition Matrix}

Consider energy ``bins'' $u$ and $l$, with $E_u > E_l$.
Suppose bin $l$ consists of many subbins $\alpha=1,...,N_\alpha$,
each of width $\Delta E_\alpha = \Delta E_l/N_\alpha$,
and bin $u$ consists of many subbinss $\beta=1,...,N_\beta$, each
with width $\Delta E_\beta = \Delta E_u/N_\beta$.
The upward Einstein $A$ coefficient from subbin $\alpha$ to subbin $\beta$
is
\beq
A_{\beta\alpha} = \frac{8\pi\nu_{\alpha\beta}^2}{hc^2}\Delta E_\beta
C_{\abs}(\nu_{\alpha\beta})~~~,
\eeq
and the spontaneous decay rate from subbin $\beta$ to subbin $\alpha$ is
\beq
A_{\alpha\beta} = \frac{(dN/dE)_\alpha}{(dN/dE)_\beta}
\frac{8\pi\nu_{\alpha\beta}^2}{hc^2} \Delta E_\alpha 
C_{\abs}(\nu_{\alpha\beta})~~~,
\label{eq:A_alphabeta}
\eeq
where $\nu_{\alpha\beta}\equiv (E_\beta-E_\alpha)/h$,
and $dN/dE$ is the vibrational density of states.
We now assume that $(dN/dE)_\alpha\approx (dN/dE)_l$, and
$(dN/dE)_\beta\approx (dN/dE)_u$.

To obtain the effective bin-to-bin transition rate $\bT_{ul}$, we
equate the total rate of energy absorption to $\bT_{ul}(E_u-E_l)$.
Thus, averaged over subbins $\alpha$, 
the transition rate from $l\rightarrow u$ is
\begin{eqnarray}
\bT_{ul}
&=&
	\frac{1}{E_u-E_l}
	\frac{c}{N_\alpha}
	\sum_\alpha
	\sum_\beta C_{\abs}(\nu) 
	u_E \Delta E_\beta
\\
&\rightarrow& 
	\frac{c}{\Delta E_l}
	\frac{1}{E_u-E_l}
	\int_{E_{l,\min}}^{E_{l,\max}} dx
	\int_{E_{u,\min}}^{E_{u,\max}} dy
	\left[ C_{\abs}(E) u_E \right]_{E=y-x} ~~~.
\end{eqnarray}
The integrand depends only on $E=y-x$.
One can show that
\beq
\bT_{ul} = 
\frac{c\Delta E_u}{E_u-E_l} 
\int_{W_1}^{W_4} dE\, G_{ul}(E) C_{\abs}(E) u_E ~~~,
\eeq
where
$W_1$, $W_4$, and $G_{ul}(E)$ are defined in equations 
(\ref{eq:G_ul}-\ref{eq:W_4}).

Similarly, averaged over subbins $\beta$, the 
energy-weighted transition rate 
from $u\rightarrow l$ is
\begin{eqnarray}
\bT_{lu} &=& \frac{1}{N_\beta}\frac{1}{E_u-E_l}
	\sum_\beta
	\sum_\alpha A_{\alpha\beta}(E_\beta-E_\alpha)
	\\
&=& \frac{(dN/dE)_l}{(dN/dE)_u}
	\frac{8\pi}{h^3c^2}
	\frac{\Delta E_\beta}{E_u-E_l}
	\sum_\beta
	\sum_\alpha E^3 C_{\abs}(E)
	\\ 
&\rightarrow& 
	\frac{(dN/dE)_l}{(dN/dE)_u}
	\frac{8\pi}{h^3c^2} 
	\frac{\Delta E_l}{\Delta E_u}
	\frac{1}{E_u-E_l}
	\int_{W_1}^{W_4} dE \, G_{ul}(E) E^3 C_{\abs}(E)\\
&=& 
	\frac{g_l}{g_u} 
	\frac{8\pi}{h^3c^2}
	\frac{1}{E_u-E_l}
	\int_{W_1}^{W_4} dE \, G_{ul}(E) E^3 C_{\abs}(E)~~~.
\end{eqnarray}
In the case of transitions to adjacent bins, we augment
$\bT_{u,u-1}$ by the rate of energy absorption in ``intrabin''
transitions divided by $E_u-E_{u-1}$, and $\bT_{u-1,u}$ by
the power radiated in intrabin transition, divided by $E_u-E_{u-1}$
(see eq.\ \ref{eq:Tu,u-1}, \ref{eq:intrabin_stat}).

\section{Emission Spectrum\label{app:spectrum}}

$F_\nu$, the power radiated per grain per unit solid angle
per unit frequency,
can be evaluated by noting from eq.\ (\ref{eq:T_lu}) that
the contribution from transitions $u\rightarrow l$ is just
\beq
\delta F_\nu \,d\nu = P_u 
\frac{g_l}{g_u}\frac{2}{h^3c^2} G_{ul}(E) E^3 C_\abs(E)
\left[ 1 + \frac{\lambda^3}{8\pi}u_E\right] dE~~~.
\label{eq:dfnudnu_1}
\eeq
If the width of bin $u$ exceeds $h\nu$, then there is an additional
contribution from intrabin 
transitions originating in
the fraction $(1-h\nu/\Delta E)$ of the subbins which are
more than $h\nu$ above $E_{u,\min}$.
Using eq.\ (\ref{eq:A_alphabeta}) with $(dN/dE)_\alpha\approx(dN/dE)_\beta$,
the intrabin contribution to $F_\nu d\nu$ is
\begin{eqnarray}
\delta F_\nu \,d\nu &=& P_u \left( 1- \frac{h\nu}{\Delta E_u}\right) 
\frac{E}{4\pi}
\frac{dA}{dE} 
\left[1+\frac{\lambda^3}{8\pi}u_E\right] dE
\\
&=&
P_u \left( 1 - \frac{h\nu}{\Delta E_u}\right) 
\frac{E}{4\pi} \,\frac{8\pi\nu^2}{hc^2}
C_\abs(E) 
\left[1+\frac{\lambda^3}{8\pi}u_E\right] dE ~~~,
\label{eq:dfnudnu_2}
\end{eqnarray}
where $dA/dE$ is obtained by dividing eq.\,(\ref{eq:A_alphabeta}) by
$\Delta E_\alpha$.
With $E=h\nu$, it is straightforward to use 
eq.\ (\ref{eq:dfnudnu_1}-\ref{eq:dfnudnu_2}) to obtain
eq.\ (\ref{eq:nuFnu}).

\section{Lemmas}
From the definition (\ref{eq:G_ul}) 
of $G_{ul}(h\nu)$ one can show that
\beq
\sum_{l=0}^{u-1} \Delta E_u \Delta E_l \, G_{ul}(h\nu)
=\min\left( h\nu,\Delta E_u \right) ~~~{\rm for}~ 
h\nu < E_{u,\min}-E_{1,\min} ~~~,
\label{eq:sum_del_Gul}
\eeq
\beq
\sum_{l=0}^{u-1} \Delta E_l (E_u-E_l) G_{ul}(E) \approx E~~~,
\label{eq:sum_Eul}
\eeq
which are used to obtain eq.\,(\ref{eq:T^cont}) and (\ref{eq:tau_rad_approx}).


\end{document}